\documentclass[10pt,twocolumn,letterpaper,dvipsnames,svgnames,x11names]{article}

\usepackage[pagenumbers]{cvpr} %

\usepackage{enumerate}
\usepackage{wrapfig}
\usepackage{multirow}
\usepackage{bbm}
\usepackage{bm}
\usepackage{booktabs}

\usepackage{pifont}
\usepackage[flushleft]{threeparttable}
\usepackage{array}
\usepackage{bibunits}

\allowdisplaybreaks[4]

\newcommand{\parsection}[1]{\noindent\textbf{#1.}~}
\let\oldparagraph\paragraph
\renewcommand{\paragraph}[1]{\oldparagraph{#1.}}
\newcommand{\tick}{{\color{Green}\ding{51}}}
\newcommand{\cross}{{\color{Red}\ding{55}}}

\newif\ifreview 
\newif\ifarxiv \newcommand{\arxiv}{\arxivtrue}
\newif\ifcamera 
\newif\ifrebuttal

\usepackage{xr-hyper}

\makeatletter
\newcommand*{\addFileDependency}[1]{%
\typeout{(#1)}%
\@addtofilelist{#1}
\IfFileExists{#1}{}{\typeout{No file #1.}}
}\makeatother

\makeatletter
\def\cref@getref#1#2{%
  \expandafter\let\expandafter#2\csname r@#1@cref\endcsname%
  \expandafter\expandafter\expandafter\def%
    \expandafter\expandafter\expandafter#2%
    \expandafter\expandafter\expandafter{%
      \expandafter\@firstoffive#2}}%
\def\cpageref@getref#1#2{%
  \expandafter\let\expandafter#2\csname r@#1@cref\endcsname%
  \expandafter\expandafter\expandafter\def%
    \expandafter\expandafter\expandafter#2%
    \expandafter\expandafter\expandafter{%
      \expandafter\@secondoffive#2}}%
      
\AtBeginDocument{%
   \def\label@noarg#1{%
    \cref@old@label{#1}%
    \@bsphack%
    \edef\@tempa{{page}{\the\c@page}}%
    \setcounter{page}{1}%
    \edef\@tempb{\thepage}%
    \expandafter\setcounter\@tempa%
    \cref@constructprefix{page}{\cref@result}%
    \protected@write\@auxout{}%
      {\string\newlabel{#1@cref}{{\cref@currentlabel}%
      {[\@tempb][\arabic{page}][\cref@result]\thepage}{}{}{}}}%
    \@esphack}%
  \def\label@optarg[#1]#2{%
    \cref@old@label{#2}%
    \@bsphack%
    \edef\@tempa{{page}{\the\c@page}}%
    \setcounter{page}{1}%
    \edef\@tempb{\thepage}%
    \expandafter\setcounter\@tempa%
    \cref@constructprefix{page}{\cref@result}%
    \protected@edef\cref@currentlabel{%
      \expandafter\cref@override@label@type%
        \cref@currentlabel\@nil{#1}}%
    \protected@write\@auxout{}%
      {\string\newlabel{#2@cref}{{\cref@currentlabel}%
      {[\@tempb][\arabic{page}][\cref@result]\thepage}{}{}{}}}%
    \@esphack}%
    }

\arxiv

\definecolor{cvprblue}{rgb}{0.21,0.49,0.74}
\usepackage[pagebackref,breaklinks,colorlinks,bookmarksdepth=3,bookmarksopen=true,bookmarksopenlevel=2,bookmarksnumbered=true]{hyperref}

\title{Make-It-Animatable: An Efficient Framework for Authoring Animation-Ready 3D Characters}

\author{
Zhiyang Guo$^{1,2}$\footnotemark[1] \and
Jinxu Xiang$^{2}$ \and
Kai Ma$^{2}$ \and
Wengang Zhou$^{1}$ \and
Houqiang Li$^{1}$ \and
Ran Zhang$^{2}$\footnotemark[2]
\\
$^{1}$CAS Key Laboratory of Technology in GIPAS, EEIS Department,\\
University of Science and Technology of China \\
$^{2}$Tencent PCG \\
{\tt\small
guozhiyang@mail.ustc.edu.cn, kylekma@tencent.com, \{zhwg, lihq\}@ustc.edu.cn,}\\
{\tt\small
\{jinxuxiang, ranorizhang\}@global.tencent.com
}
\vspace{-20pt}
}

\begin{document}

\twocolumn[
{
\maketitle
\vspace{-10pt}
\begin{center}
  \centering
  \includegraphics[width=\linewidth, trim = 0cm 0cm 0cm 0cm, clip]{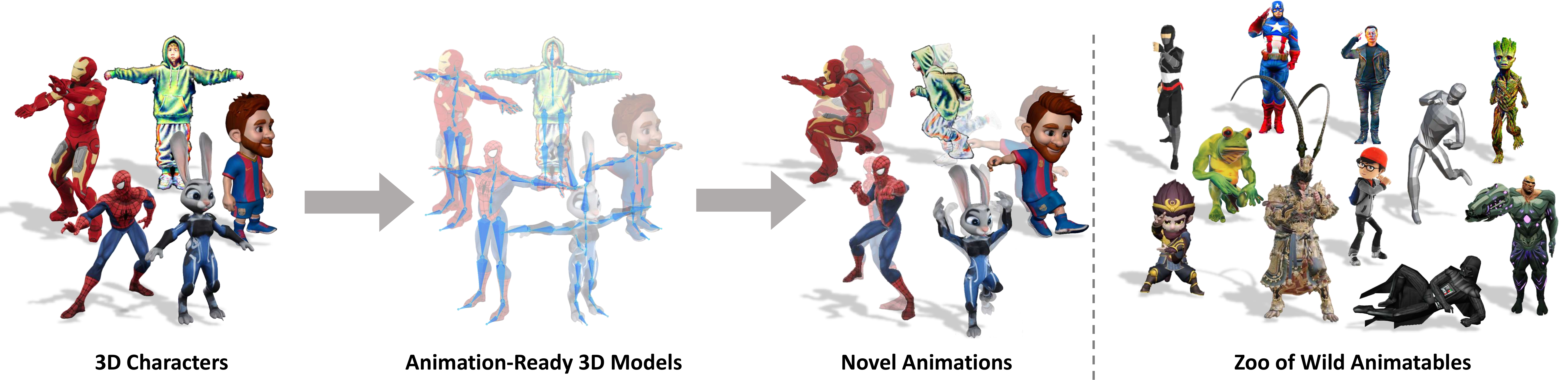}
\end{center}
\vspace{-10pt}
\captionof{figure}{Given a 3D character represented by mesh or 3D Gaussian Splats with arbitrary pose and shape, our framework can produce high-quality results of rigging, skinning, and pose resetting for it within one second. The output 3D model is fully animatable with a fine-grained skeleton and optional bone topology of extra body structures.}
\label{fig:teaser}
\vspace{15pt}
}
]

{
  \renewcommand{\thefootnote}%
    {\fnsymbol{footnote}}
  \footnotetext[1]{Work is done during internship at Tencent.}
  \footnotetext[2]{Corresponding author.}
}

\begin{abstract}

\vspace{-1mm}
3D characters are essential to modern creative industries, but making them animatable often demands extensive manual work in tasks like rigging and skinning. 
Existing automatic rigging tools face several limitations, including the necessity for manual annotations, rigid skeleton topologies, and limited generalization across diverse shapes and poses. 
An alternative approach is to generate animatable avatars pre-bound to a rigged template mesh. 
However, this method often lacks flexibility and is typically limited to realistic human shapes. 
To address these issues, we present \textbf{Make-It-Animatable}, a novel data-driven method to make any 3D humanoid model ready for character animation in less than one second, regardless of its shapes and poses. 
Our unified framework generates high-quality blend weights, bones, and pose transformations. By incorporating a particle-based shape autoencoder, our approach supports various 3D representations, including meshes and 3D Gaussian splats. Additionally, we employ a coarse-to-fine representation and a structure-aware modeling strategy to ensure both accuracy and robustness,
even for characters with non-standard skeleton structures.
We conducted extensive experiments to validate our framework's effectiveness. Compared to existing methods, our approach demonstrates significant improvements in both quality and speed.
More demos and code are available at \url{https://jasongzy.github.io/Make-It-Animatable/}.
\end{abstract}

\vspace{-3mm}
\section{Introduction}
\label{sec:intro}

3D characters, as the principal subjects of the digital world, play an essential role in various fields of modern creative industries, \eg, video games, 3D animations, films, mixed reality, \etc. 
Bringing 3D characters to life requires making them animatable, a crucial yet labor-intensive task that involves substantial effort in rigging and skinning.
Additionally, animation-ready character models must be positioned in predefined rest poses to support effective retargeting and deformation.
Traditionally, rigging and skinning processes are either manually executed by artists, demanding considerable time and effort for each task, or managed using existing automatic rigging tools that often lack the robustness and generalizability to accommodate diverse character types and animation requirements.
Recent advancements have sought to address these challenges by employing parameterized templates to directly generate animatable characters as meshes or 3D Gaussian splats.
However, these methods are frequently constrained by their dependence on realistic humanoid models and lack the flexibility to handle non-standard poses or shapes, limiting their applicability to the dynamic and varied 3D character designs prevalent in modern applications.

In this work, we present a comprehensive data-driven framework that enables instant rigging and skinning of 3D characters, regardless of their initial shapes, poses, or structural complexities. 
Our solution is exceptionally fast, processing each character in approximately one second while generating high-quality, animation-ready models with accurate bones, blend weights, and rest pose transformations.
Unlike other automatic rigging approaches, our system adeptly handles challenging cases, including exaggerated characters with unconventional head and body proportions, as well as non-standard poses featuring additional bone structures such as hands, ears, and tails.

\cref{tab:function}
compares our method with existing approaches across key features, including the ability to handle both mesh and 3D Gaussian splatting inputs, flexibility in adapting to various poses, and support for advanced animation features such as hand and finger articulation.

Unlike commercial auto-rigging tools like ~\cite{mixamo,anythingworld}, our model is designed to work with any predefined skeleton structure, providing greater control over joint movements and significantly improving rigging speed and flexibility.
Our framework's rapid and precise rigging and skinning of diverse 3D characters unlocks new possibilities for dynamic character animations. This capability is particularly beneficial in applications requiring swift responses and high customization, such as virtual reality, gaming, and real-time simulations. Additionally, it serves as a valuable enhancement to existing static 3D model generation systems, enabling the creation of dynamic 3D models.

\begin{table}[t]
    \centering
    \resizebox{1.0\linewidth}{!}{
    \begin{threeparttable}
    \begin{tabular}{c|l|c|c|c|c|c|c|c}
        \toprule
        \multirow{2}*{\textbf{Categories}} & \multirow{2}*{\textbf{Methods}} & \multirow{2}*{\textbf{Mesh}} & \multirow{2}*{\textbf{3DGS}} & \multirow{2}*{\shortstack{\textbf{Template}\\\textbf{Free}}} & \multirow{2}*{\shortstack{\textbf{Alterable}\\\textbf{Skeleton}}} & \multirow{2}*{\shortstack{\textbf{Pose to}\\\textbf{Rest}}}& \multirow{2}*{\shortstack{\textbf{Hand}\\\textbf{Animation}}} & \multirow{2}*{\shortstack{\textbf{Rigging}\\\textbf{Time}}}\tnote{3}  \\
        &  &  &  &  &  &  &  & \\
        \midrule
        \multirow{2}*{\footnotesize{\shortstack{Text/Image to\\Animatable}}}
        & Meshy\tnote{$\dagger$}~~\tnote{1}~~\cite{meshy} & \tick & \cross & \tick & \cross & \tick & \cross & $\sim$~3~min \\
        & Tripo\tnote{$\dagger$}~~\cite{tripo} & \tick & \cross & \tick & \cross & \tick & \cross & $\sim$~3~min \\
        \midrule
        \multirow{4}*{\footnotesize{\shortstack{Auto Rigging}}}
        & Mixamo\tnote{$\dagger$}~~\tnote{1}~~\cite{mixamo} & \tick & \cross & \tick & \cross & \tick\tnote{2} & \tick & $\sim$~2~min \\
        & {{\footnotesize Anything World}}\tnote{$\dagger$}~~\cite{anythingworld} & \tick & \cross & \tick & \tick & \tick\tnote{2} & \tick & $\sim$~4~min \\
        & RigNet~\cite{xu2020rignet} & \tick & \cross & \tick & \tick & \cross & \cross & $\sim$~10~min \\
        & TARig~\cite{ma2023tarig} & \tick & \cross & \tick & \tick & \cross & \cross & $\sim$~0.6~min \\
        \midrule
        \multirow{2}*{\footnotesize{\shortstack{Template-based\\Animatable}}}
        & TADA~\cite{liao2024tada} & \tick & \cross & \cross & \cross & \textbf{--} & \tick & \textbf{--} \\
        & {{\footnotesize HumanGaussian}}~\cite{liu2024humangaussian} & \cross & \tick & \cross & \cross & \textbf{--} & \cross & \textbf{--} \\
        \midrule
        & Ours & \tick & \tick & \tick & \tick & \tick & \tick & $\sim$~0.5~s \\
         \bottomrule
    \end{tabular}

    \smallskip%
    \begin{tablenotes}
        \item[$\dagger$] Commercial software.
        \item[1] Meshy and Mixamo are semi-automatic rigging methods, requiring manual annotation of joint positions.
        \item[2] Mixamo and Anything World only support poses close to T- or A-pose, while our methods allow for arbitrary poses.
        \item[3] Tested on a typical input mesh with 8k vertices. 
    \end{tablenotes}

    \end{threeparttable}
    }
    \vspace{-5pt}
    \caption{
    \textbf{Comparison between our method and existing approaches in terms of key features.}
    }
    \label{tab:function}
    \vspace{-10pt}
\end{table}

\section{Related Works}
\label{sec:related}

\subsection{Modeling Dynamics in 3D Vision and Graphics}

Modeling dynamics is a fundamental task in computer vision and computer graphics.
It involves both the spatial representation of 3D models and the formulation of their temporal behaviors.
Recent approaches in this area can be categorized into mesh-based, mesh-free, reduced-order, and proxy-based methods.

Mesh-based elasticity models have been a cornerstone in modeling dynamics within computer graphics~\cite{terzopoulos1987elastically}.
One of the most commonly employed variants is the linear tetrahedral finite element method~\cite{Sifakis2012FEM, cutler2002procedural,bargteil2007finite,bouaziz2014projective},
which requires a robust surface-to-tetrahedron algorithm like TetWild~\cite{hu2018tetrahedral}.

However, both the linear finite element method and tetrahedral generation are sensitive to the geometry of the model, necessitating special handling for intricate shapes, such as thin features~\cite{Baraff1998cloth} and rod-like geometries~\cite{Bergou2008rods}.
For complex geometries, including non-manifold shapes, mesh-free methods, first introduced by~\cite{Desbrun1995implicit}, provide an effective solution for addressing the challenges of mesh-based dynamic modeling.
For example, smoothed-particle hydrodynamics (SPH)~\cite{monaghan2005smoothed,peer2018implicit} and material point method (MPM)~\cite{stomakhin2013material,Jiang2016MPM,muller2004point}, which use particles as the main spatial representation, overcomes the difficulties of modeling different materials and complex dynamics in mesh-based methods, yielding high-quality results.

One major drawback of particle-based methods is their high computational complexity, resulting from the dense sampling required. To address this, researchers introduced reduced-order dynamics~\cite{barbivc2011real,barbivc2005real,barbic2007real}, which decrease the degrees of freedom by projecting motions onto a limited set of deformation bases. This approach significantly accelerates computations and is applicable to both mesh-based and mesh-free methods. Recently, methods such as CROM~\cite{chen2022crom} and LiCROM~\cite{chang2023licrom} have advanced continuous reduced-order modeling using neural networks.

Reduced-order dynamics simplify temporal behaviors by reducing their dimensionality. Similarly, spatial representations can be simplified through proxy-based methods, which employ low-dimensional proxies to efficiently compute dynamics and subsequently apply deformations to the original geometry. Notable proxy-based techniques include Free-Form Deformation (FFD)~\cite{sederberg1986free} and cage-based methods~\cite{stroter2024survey,ju2008reusable,coros2012deformable,kim2014interactive}. These approaches utilize simplified spatial representations to expedite dynamic calculations, which are then used to deform the detailed geometry. 

The most widely used proxy-based method is Linear Blend Skinning (LBS).
It deforms a mesh by blending transformations from an underlying skeletal structure, allowing for efficient and intuitive animation of complex models. This technique was introduced by Magnenat-Thalmann et al.~\cite{magnenat1988joint} and further developed by Lewis et al.
~\cite{Lewis2000pose}.
Recent advancements with neural networks have further enhanced LBS, making it suitable for geometry-agnostic simulation~\cite{Modi2024simplicits} and enabling more natural deformations through neural blend shapes~\cite{li2021learning}.

\begin{figure*}[t]
      \centering
      \includegraphics[width=0.95\linewidth]{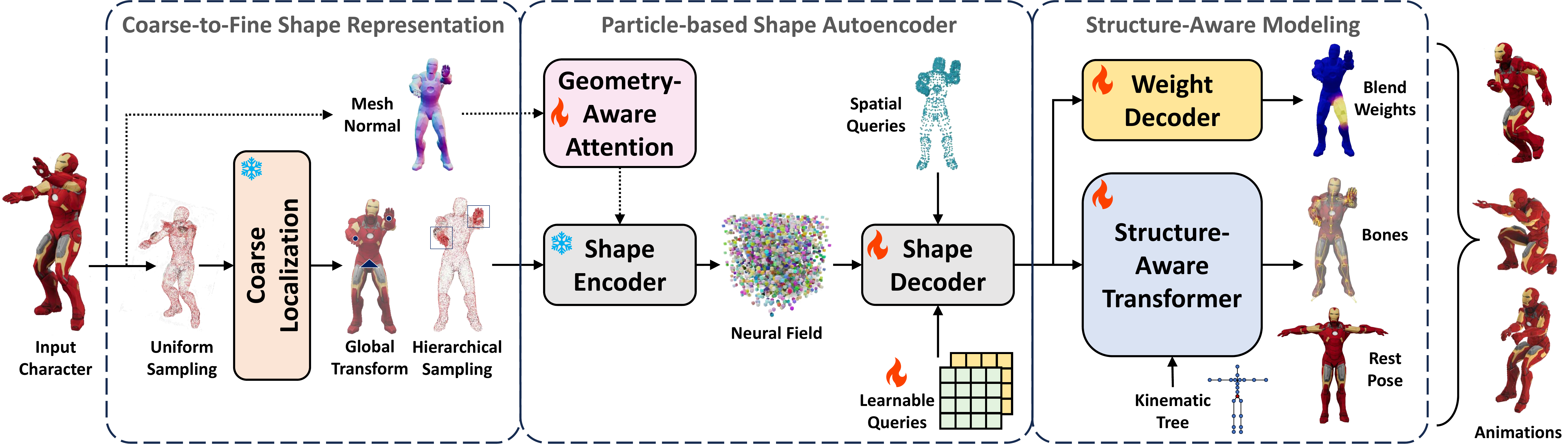}
      \caption{
      \textbf{Pipeline of the proposed framework.}
      Given an input 3D character, we produce high-quality blend weights, bones, and pose-to-rest transformation for it, so that any animation is within easy reach.
      First, we coarsely localize the joints with a pre-trained lite version of this framework, which helps enable a finer shape representation.
      Then the shape is encoded into a neural field with a particle-based autoencoder. The decoding process involves spatial and learnable queries for different animation assets.
      Finally, the structure-aware modeling of bones is proposed to better align the predictions with skeleton topology priors.
      }
      \label{fig:pipeline}
      \vspace{-10pt}
\end{figure*}

\subsection{Authoring Animation-Ready 3D Models}
In the 3D animation and gaming industry, Linear Blend Skinning (LBS) is the standard for character animation. Creating animation-ready 3D models with LBS involves two key processes: constructing rigs and assigning skinning weights to the input models. Traditionally, these tasks are performed manually in 3D modeling software like Autodesk Maya, where artists place rigs and paint skinning weights to achieve the desired deformations. 

Pinocchio~\cite{baran2007automatic} is a pioneering system for automatic rigging and skinning of 3D characters.
It automates the embedding of a skeleton into a 3D mesh and assigns skinning weights based on vertex proximity to bones, enabling smooth and natural deformations during animation.
Recent advancements in automatic rigging have leveraged deep learning techniques to enhance flexibility and accuracy.
RigNet~\cite{xu2020rignet} utilizes neural networks trained on extensive datasets of animated characters to predict rigging parameters, enabling the generation of custom rigs suitable for diverse character models.
Similarly, TARig~\cite{ma2023tarig} employs a template-aware approach, combining a humanoid skeleton with a shared graph neural network backbone to ensure precise joint positioning and skin weight estimation, thereby improving rigging efficiency.  
To enhance skinning quality, Li et al.~\cite{li2021learning} proposed neural blend shapes in addition to predicting rigs, achieving better deformation compared to static skinning weights.
Another advancement in automatic rigging involves relaxing the requirement for the input pose to be in a standard position, such as A-pose or T-pose; Theisel et al~.\cite{Theisel2021bipedal} introduced a method that generates rigs for characters in arbitrary poses.
Nonetheless, the inadequate quality, the limited robustness against complex inputs, and the unsatisfying time cost, all hinder the practical application of these auto-rigging methods.

An alternative approach to creating animation-ready models involves generating 3D models that are pre-bound to a rigged template. For instance, TADA~\cite{liao2024tada} produces textured human meshes by deforming a SMPL-X~\cite{pavlakos2019smplx} template through score distillation sampling (SDS) optimization~\cite{poole2022dreamfusion}. 
Similarly, HumanGaussian~\cite{liu2024humangaussian} generates human models in Gaussian splats from a SMPL~\cite{loper2015smpl} mesh using a structure-aware SDS algorithm.
Additionally, DreamWaltz-G~\cite{huang2024dreamwaltzg} employs a skeleton-guided distillation method combined with a hybrid 3D Gaussian avatar representation to achieve realistic generation and expressive animation.
Although the aforementioned works have introduced various strategies to improve the shape diversity of generation, they remain constrained by the preset body ratio of SMPL mesh and are limited to avatars resembling realistic humans.
Moreover, the fixed template skeleton topology also prevents their extension to non-standard bone structures.

\section{Method}
\label{sec:method}

\subsection{Preliminaries}
\label{sec:pre}

Intuitively, modeling the dynamics of an object requires a per-timestamp deformation of all the particles (\ie, vertices for meshes or splats for 3D Gaussians) that make up its geometry.
Suppose that we have an object composed of $N$ particles and a desired dynamic sequence of $T$ timestamps. The temporal deformations of all the particles can be denoted by a matrix $\bm{D} \in \mathbb{R}^{T\times N \times d}$, where $d$ is the degrees of freedom (\eg, $d=6$ for rigid transformations).
Generally, $\bm{D}$ has a lot of redundancy when representing real-world dynamics, so we would like to seek a low-rank approximation expressed by
$\bm{D}^{T\times N \times d} \approx \bm{B}_t^{T \times K \times d} \bm{W}_s^{K \times N}$,
where $K$ is the desired rank, $\bm{B}_t$ and $\bm{W}_s$ are \textit{temporal basis} and \textit{spatial weight} matrices, respectively.
The practical applications of this low-rank form of dynamics include the sparse control paradigms~\cite{huang2024scgs,lei2024mosca} and the linear blend skinning (LBS)~\cite{loper2015smpl,pavlakos2019smplx} algorithm, where $\bm{B}_t$ is interpreted as the transformations/poses of control nodes/body joints, and $\bm{W}_s$ is the blend weights of the nodes or joints. Since the motions under the same joint setting can be reused, solving $\bm{B}_t$ for humanoid characters is usually converted to finding the transformations from all poses to a predefined rest pose.

Based on this formulation, given a shape representation of a character composed of $N$ particles, all the assets necessary for animating it (named as \textit{animation assets}) are:
\textbf{1)} $K$ joints
represented by bone head and tail positions, $\bm{J} \in \mathbb{R}^{K \times 6}$, which are connected based on any predefined skeleton topology;
\textbf{2)} a 6-DoF transformation, $\bm{P} \in \mathbb{R}^{K \times 6}$, which transforms the input posed joints to a common rest pose;
\textbf{3)} the blend weights $\bm{W} \in \mathbb{R}^{K \times N}$ between any joint-particle pair.
The former two are also referred to as \textit{bone attributes}, which are the abstract representation encompassing all the input particles (instead of being tied to individual particles like the blend weights).

In the following subsections, we will elaborate on how these animation assets are obtained with our proposed unified framework, which is illustrated in \cref{fig:pipeline}. 
For the sake of clarity, we will first introduce the core part of this framework --- a shape autoencoder that describes the input geometry as compact neural latent representations.

\subsection{Particle-based Shape Autoencoder}
\label{sec:autoencoder}

\parsection{Neural field encoding}
The encoding part of our shape autoencoder is similar to the downsampled-point querying architecture of 3DShape2VecSet~\cite{zhang20233dshape2vecset}.
We start from a point cloud $\bm{X} \in \mathbb{R}^{N \times 3}$ sampled from the input character shape after re-centering and normalizing.
In order to learn a compact representation for this shape, we want to aggregate the information in this possibly large point cloud into a smaller set (size $M$) of latent feature vectors denoted by $\bm{F} \in \mathbb{R}^{M \times C}$.
To achieve this, $\bm{X}$ is first downsampled to $M$ points with farthest point sampling (FPS)~\cite{qi2017pointnet}: $\tilde{\bm{X}} = \operatorname{FPS}(\bm{X}) \in \mathbb{R}^{M \times 3}$. Then equipped with a positional encoding mapping denoted by $\operatorname{PE}: \mathbb{R}^{3} \rightarrow \mathbb{R}^{C}$, the latent features $\bm{F}$ are obtained via a cross-attention between the original and the downsampled points:
 \begin{equation}
     \bm{F} = \operatorname{CrossAttn}(\operatorname{PE}(\tilde{\bm{X}}),\operatorname{PE}(\bm{X}))
     \in \mathbb{R}^{M \times C}.
 \end{equation}
In this way, the network can adaptively encode the spatial information as a neural field, instead of representing the actual spatial position explicitly.
In practice, the shape encoder pretrained on ShapeNet~\cite{chang2015shapenet} already has sufficient capacity for low-level geometry perception. Therefore, we freeze the parameters of the shape encoder during the training of our framework.

\parsection{Geometry-aware attention}
One of the benefits of our particle-based autoencoder is the support for various shape representations (as long as point sampling is possible on them). Nevertheless, for surface representations like mesh, much geometric information about the shape is lost during the sampling.
This can lead to blend weight corruption problems that semantically distant but spatially close points may incorrectly share similar weight values.
To address this issue, a non-intrusive way of injecting geometric awareness into the shape representation is proposed.
Specifically, we extract the per-point normal values from the input mesh, as they carry rich geometry information.
Then a lightweight fusing module is implemented using a simple attention layer as a branch way to the shape encoder, whose parameters are initialized so that the normal branch produces zero impact on the encoding results at the beginning.
We randomly corrupt the normal values in training to make it more robust to low-quality meshes and also compatible with inputs like 3DGS in inference.
Eventually, the attention mechanism adaptively decides what regions should care about the normals and whether the given normal values carry useful cues.

\parsection{Spatially continuous decoding for blend weights}
The neural field represented by latent features $\bm{F}$ can be directly queried with coordinates to obtain spatially continuous attributes,
\ie, the blend weights.
Our shape decoder completes decoding in a flexible and learnable way by applying attention between the queries and the neural fields. Formally, given the $N_q$-point spatial queries (\eg, all vertices of a mesh) denoted by $\bm{Q}_w \in \mathbb{R}^{N_q \times 3}$, the corresponding shape-aware embeddings $\bm{F}_w$ is derived by
\begin{equation}
\hspace{-2mm}
    \bm{F}_w = \operatorname{CrossAttn}(\operatorname{PE}(\bm{Q}_w), \operatorname{SelfAttn}(\bm{F}))
    \in \mathbb{R}^{N_q \times C}.
\label{eq:decode_bw}
\end{equation}
Then using an MLP-based lightweight decoding head $\Theta_w$, we can finally get the blend weights $\bm{W} = \Theta_w(\bm{F}_w) \in \mathbb{R}^{K \times N_q}$ for all the query coordinates.

\begin{figure*}[t]
      \centering
      \includegraphics[width=0.9\linewidth]{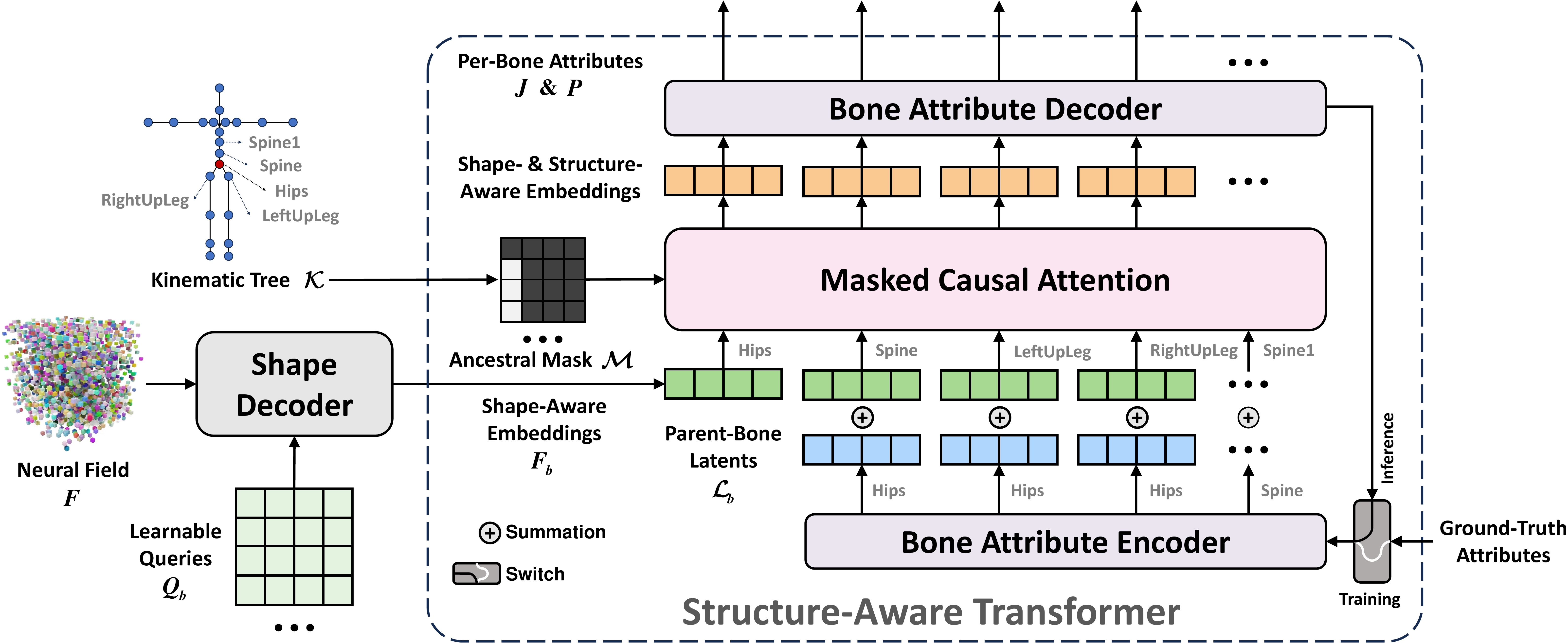}
      \caption{
      \textbf{Pipeline of the proposed structure-aware transformer.}
      The per-bone shape-aware embedding is first added with its parent bone's latent, which is encoded from the autoregressive outputs (in inference) or the ground-truth values (in training). The summation is then fused with the ancestral bones' features via the masked causal attention. Eventually, bone attributes are decoded from the output shape- and structure-aware embeddings.
      In inference, the whole process follows the paradigm of next-child-bone prediction.
      }
      \label{fig:transformer}
      \vspace{-10pt}
\end{figure*}

\parsection{Learnable discrete decoding for bone attributes}
When it comes to discrete per-bone attributes like joints' positions and poses, the decoding process becomes more tricky.
While following the same querying pattern~\cite{loper2015smpl,ma2023tarig} of blend weights and predicting a joint regressor is possible, it’s inefficient for dense inputs and makes bone attributes sensitive to the spatial queries.
Therefore, we adopt another way that involves learnable semantic queries.
Specifically, for an attribute of the predefined $K$ bones, we assign $K$ learnable query embeddings denoted by $\bm{Q}_b \in \mathbb{R}^{K \times C}$. Then we use attention layers to integrate the shape features, similar to \cref{eq:decode_bw}:
\begin{equation}
\vspace{-1mm}
    \bm{F}_b = \operatorname{CrossAttn}(\bm{Q}_b, \operatorname{SelfAttn}(\bm{F}))
    \in \mathbb{R}^{K \times C}.
\end{equation}
Now the bone-wise embeddings $\bm{F}_b$ contain both the global geometry information and bone semantic cues, \ie, which region should a certain bone attend to.
Then two individual bone attribute decoders are employed to regress the desired per-bone head-and-tail positions $\bm{J} \in \mathbb{R}^{K \times 6}$ and pose-to-rest transformation $\bm{P} \in \mathbb{R}^{K \times 6}$, respectively.
We found in practice that the rigid transformation represented by dual quaternions presents better behaviors in optimization than a vanilla 6-DoF movement. Therefore, the pose decoder actually outputs 8D dual quaternions $\bm{P}_{dq} \in \mathbb{R}^{K \times 8}$ that can be trivially converted to a rigid transformation matrix.

\subsection{Coarse-to-Fine Shape Representation}
\label{sec:sampling}

With the shape autoencoder, our model can produce promising blend weight predictions, but joint outputs, particularly for fine-grained regions like the hands, still face convergence challenges. 
We attribute these issues to the ambiguity of the input points and view the aforementioned pipeline as a coarse stage that provides rough but valuable localization information about the input character. 
To enhance the shape representation, we adopt two strategies: canonical transformation and hierarchical sampling. 
The canonical transformation resolves pose ambiguity by aligning the input shape to a common orientation, while the hierarchical sampling method ensures higher sampling density in key regions like the hands to improve accuracy without increasing computational cost.
Further implementation details are provided in the supplementary material.

\subsection{Structure-Aware Modeling of Bones}
\label{sec:kinematics}

With a learnable query assigned to each bone, we can effectively predict the bone-wise attributes. Although those queries adaptively attend to different regions of the input shape, they lack awareness of the skeleton topology, leading to imperfections in predictions, especially for deep-level bones in a kinematic tree.
Therefore, we propose to model the bones in a structure-aware way, including additional designs for both network architecture and loss functions.

\begin{figure*}[t]
      \centering
      \includegraphics[width=0.95\linewidth]{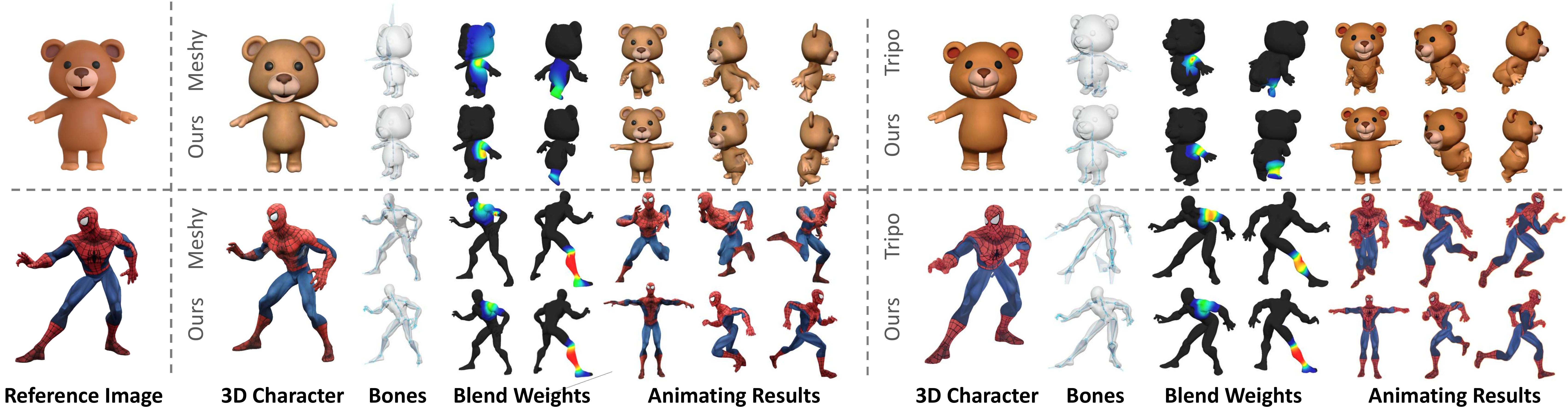}
      \caption{
      \textbf{Comparison with Meshy~\cite{meshy} and Tripo~\cite{tripo}.}
      We feed them the same image as reference and compare the performance based on their generated 3D models respectively.
      The blend weights of two joints, \ie, Left Shoulder and Right Leg, are visualized.
      Given that these baselines can only apply preset motions and their rest-pose models cannot be exported, we apply a similar ``running'' sequence to all the methods for fair comparison.
      The T-pose models predicted by our method are also included as the front-view animating results.
      }
      \label{fig:meshy_tripo}
      \vspace{-10pt}
\end{figure*}

\parsection{Next-child-bone prediction via causal attention}
Without structure awareness of the skeleton, the model tends to produce independent homogeneous predictions.
For example, the pose of a finger bone largely depends on the poses of its ancestral bones like hand and arm.
If the pose of the direct parent is known, predicting the finger's pose becomes easier, as it only requires accounting for the subtle transformation relative to its parent.
To this end, inspired by the decoder-only next-token prediction paradigm prevalently adopted in LLMs~\cite{vaswani2017attention,brown2020language}, we develop a structure-aware transformer based on a novel \textit{next-child-bone prediction} architecture, as depicted in \cref{fig:transformer}.

Let us start from the bone-wise shape-aware embeddings, $\bm{F}_b = \{\bm{f}_1,\bm{f}_2,\dots,\bm{f}_{K} \in \mathbb{R}^C\}$, produced by the shape decoder (\cref{sec:autoencoder}). 
Instead of directly feeding $\bm{F}_b$ into a bone attribute decoder, we add several masked causal attention layers to enhance them. Each bone embedding, $\bm{f}_i$,~$i \in [1, K]$, is regarded as an individual query token.
Meanwhile, we use a bone attribute encoder to map the actual attribute values (joint positions or poses) into some same-shaped bone-wise latents
$\bm{\mathcal{L}}_b = \{\bm{l}_1,\bm{l}_2,\dots,\bm{l}_{K} \in \mathbb{R}^C\}$.
Then given a kinematic tree denoted by $\mathcal{K}$, to let $\bm{f}_i$ learn about the context information from its parent bone, we add the corresponding parent-bone latent $\bm{l}_j$,~$j=\operatorname{Anc}(i, \mathcal{K})(1)$ to it, where $\operatorname{Anc}(i,\mathcal{K})$ returns a tuple of ancestral bones for the bone $i$ recursively from the nearest up to the root node in $\mathcal{K}$, and here we only select the first element (\ie, the direct parent bone).

We then feed the summation of $\bm{f}_i$ and $\bm{l}_j$ into a masked causal attention module that fuses that per-bone latent with those of other bones.
We build an ancestral mask derived from the kinematic tree, ensuring that each bone learns from its ancestors and treats the states of its children as ``future'' information. The ancestral mask is defined by
$\mathcal{M} \in \mathbb{R}^{K\times K}$ where $\mathcal{M}_{i,j} = 1~\text{iff}~ j\in \operatorname{Anc}(i, \mathcal{K})$. Equipped with $\mathcal{M}$, the cross-attention operation outputs per-bone embeddings that have both shape and structure awareness. 
Eventually, they are sent into a bone attribute decoder to be translated back to the predicted attributes.

Similar to the causal attention in language models, our structure-aware transformer follows an autoregressive process at the inference stage. The ground-truth parent-bone attributes are replaced by the predictions produced by the decoder. It takes several progressive iterations to complete the decoding and one level in the kinematic tree is predicted in an iteration, hence the name ``next-child-bone prediction''.

\parsection{Body prior loss}
Our predictions of bone attributes are directly supervised by the ground-truth values. Since a character can have multiple possible solutions in practice, we further constrain the optimizing process using prior losses, so that the predictions follow some essential body patterns.
Specifically, three kinds of prior knowledge are leveraged as follows.
1) Bone connectivity: most bone heads have to be connected to the tails of their parent bones.
2) Bone symmetry: typically in the predefined rest pose, the left and right parts of the skeleton have to be symmetric about the spine.
3) Bone parallelism: in the rest pose, multiple bones belonging to the same limb have to share the same direction.

\subsection{Training and Inference}
\label{sec:train}

The proposed framework is trained in an end-to-end data-driven manner, supervised by the $L_1$ losses with the ground-truth blend weights, bone positions, and pose-to-rest transformations, as well as the extra body prior losses (\cref{sec:kinematics}).
As introduced in \cref{sec:sampling}, a coarse-to-fine training strategy is adopted.
In the coarse stage, we uniformly sample the input shape and predict only the bone positions, applying random 3D rotations for data augmentation to enhance generalization. Then in the fine stage, we transform the input character to canonical coordinates and apply the hierarchical sampling. 

Once the training is finished, the model can take any particle-based character (in any pose with any global transformation) and infer its animation assets.
With a fixed number of neural shape latents and a learnable discrete querying strategy, our framework achieves efficient inference, with speed less affected by input particle number.
Generally, the whole feed-forwarding inference takes less than 1 second.

Although our framework requires a fixed definition of kinematic tree, and we typically choose a standard human skeleton to train it on, it can actually be extended to any predefined skeleton topology.
If the framework has already been trained with the standard skeleton, adapting it to include extra bones is as simple as fine-tuning the final layer of the weight decoder and the extra learnable queries.
In \cref{sec:exp_results}, we will present some animating results of characters with additional accessories like long ears or tails.

\section{Experiments}
\label{sec:exp}

\subsection{Experimental Settings}
\label{sec:exp_set}

\parsection{Dataset}
We utilize a collection of artist-designed 3D models from Mixamo~\cite{mixamo}, consisting of 95 high-quality characters and 2,453 diverse motion sequences. Each character is preprocessed to conform to a standard skeleton structure with $K=52$ bones, and the motion sequences contain an average of 200 frames. We allocate 95\% of the data for training and the remaining 5\% for validation. During each training iteration, a character-motion pair is randomly selected, resulting in an effective dataset size equivalent to over 40 million frames.

Furthermore, our training framework accommodates additional skeleton topologies. We used VRoid Studio~\cite{vroid} to manually create 35 distinct anime characters, each featuring accessories like animal ears and tails. These characters were preprocessed to match the Mixamo skeleton definition, with the addition of extra bones for these accessories.

We select bipedal humanoid characters from the `ModelsResource-RigNetv1' dataset~\cite{xu2019predicting} to construct a test set for quantitative comparison with existing automatic rigging algorithms. For qualitative evaluation, we gather a diverse collection of in-the-wild 3D characters from various sources, including the Objaverse dataset~\cite{deitke2023objaverse}, generative tools~\cite{meshy,tripo}, and other artist-designed models.

\begin{figure}[t]
      \vspace{-10px}
      \centering
      \includegraphics[width=0.95\linewidth]{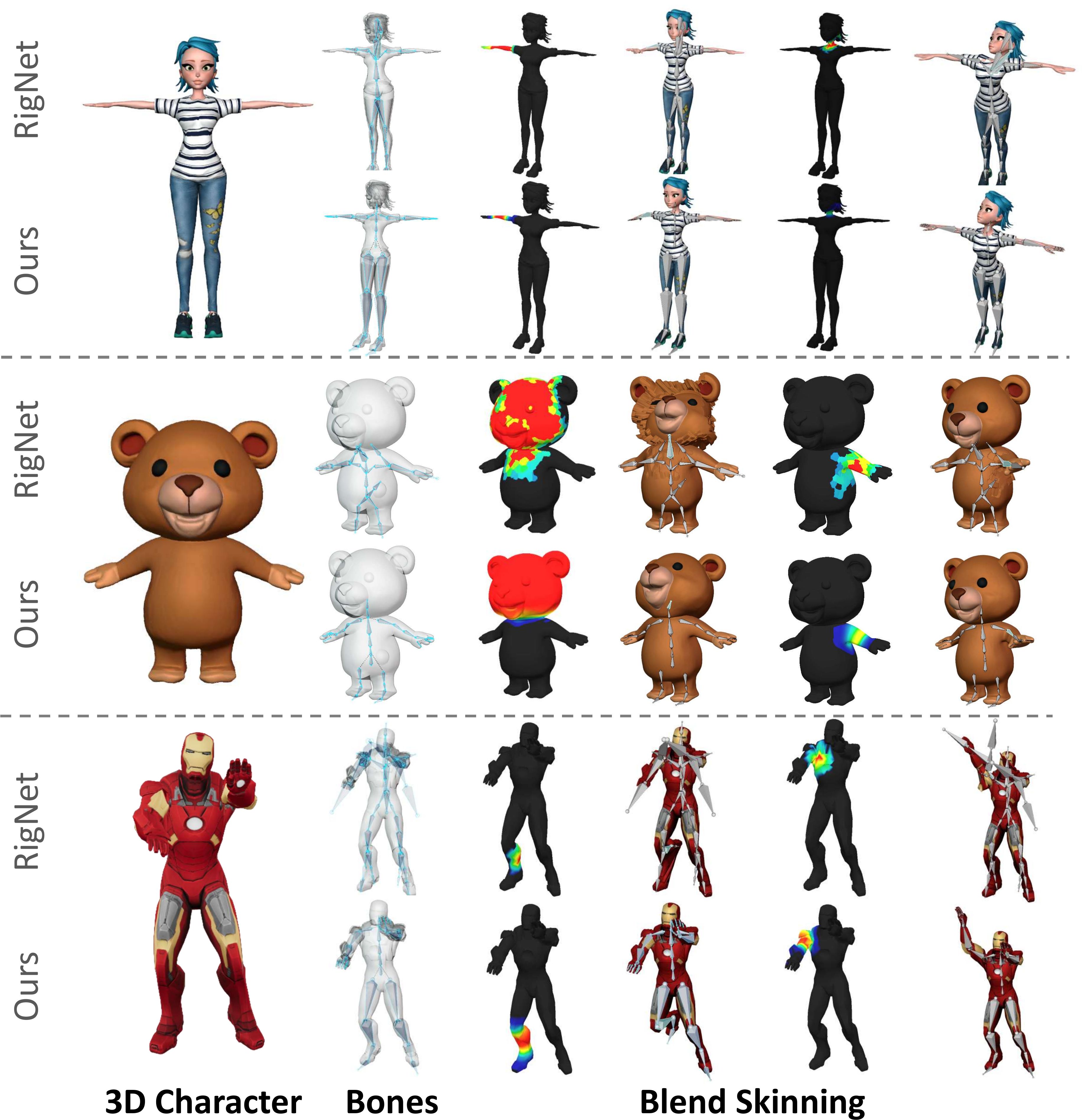}
      \vspace{-6px}
      \caption{
      \textbf{Comparison with RigNet~\cite{xu2020rignet}.}
      We visualize the blend weights of selected joints and manually deform them to assess the impact of rigging quality on skinning results.
      }
      \vspace{-10px}
      \label{fig:rignet}
\end{figure}

\parsection{Baselines}
We choose two lines of existing methods that can produce animatable characters as our baselines.
\textit{Auto-rigging methods.} These methods share the same workflow as ours, \ie, taking a 3D character as the input and outputting a rigged one with bones and blend weights.
a) Meshy~\cite{meshy}: a commercial software that generates general 3D meshes from text/image prompts, and optionally rigs the meshes of bipedal humanoid inputs. Since it does not support direct 3D inputs, we feed an image of the test 3D character into it.
b) Tripo~\cite{tripo}: another commercial software that basically has the same functions as Meshy, but differs in quality.
c) RigNet~\cite{xu2020rignet}: a data-driven model that can rig any mesh-based geometry. We feed the same test characters to both RigNet and our model.
\textit{Template-based avatar generating methods.}
By leveraging the SMPL template~\cite{loper2015smpl,pavlakos2019smplx} as a strong prior, this line of methods directly generates animatable humans.
a) TADA~\cite{liao2024tada}: mesh-based avatar generation. b) HumanGaussian~\cite{liu2024humangaussian} (HG): Gaussian-splatting-based avatar generation.
The generated avatars of these two works are rigged with their corresponding template skeletons and blend weights.

\parsection{Metrics}
In the ablations conducted on the test split of Mixamo dataset, we report the percentage error between the predicted and ground-truth animation assets.
As for the quantitative comparison with RigNet~\cite{xu2020rignet}, we employ several metrics to evaluate the quality of skeleton predictions, \ie, the IoU, Precision and Recall of bone matching, and the CD-J2J, CD-J2B, and CD-B2B (Chamfer distances between joints and bone line segments)~\cite{xu2019predicting}.
For other in-the-wild test cases without ground truth, we provide extensive qualitative comparison by visualizing the skeletons, blend weights, and animating results. Due to the different skeleton topologies, the same motion sequence is usually not applicable to other baselines. For fairness, we manually set the bones from different methods to similar poses in 3D modeling software and render the animations.

\subsection{Comparison Results}
\label{sec:exp_results}

\begin{figure}[t]
      \vspace{-10px}
      \centering
      \includegraphics[width=0.95\linewidth]{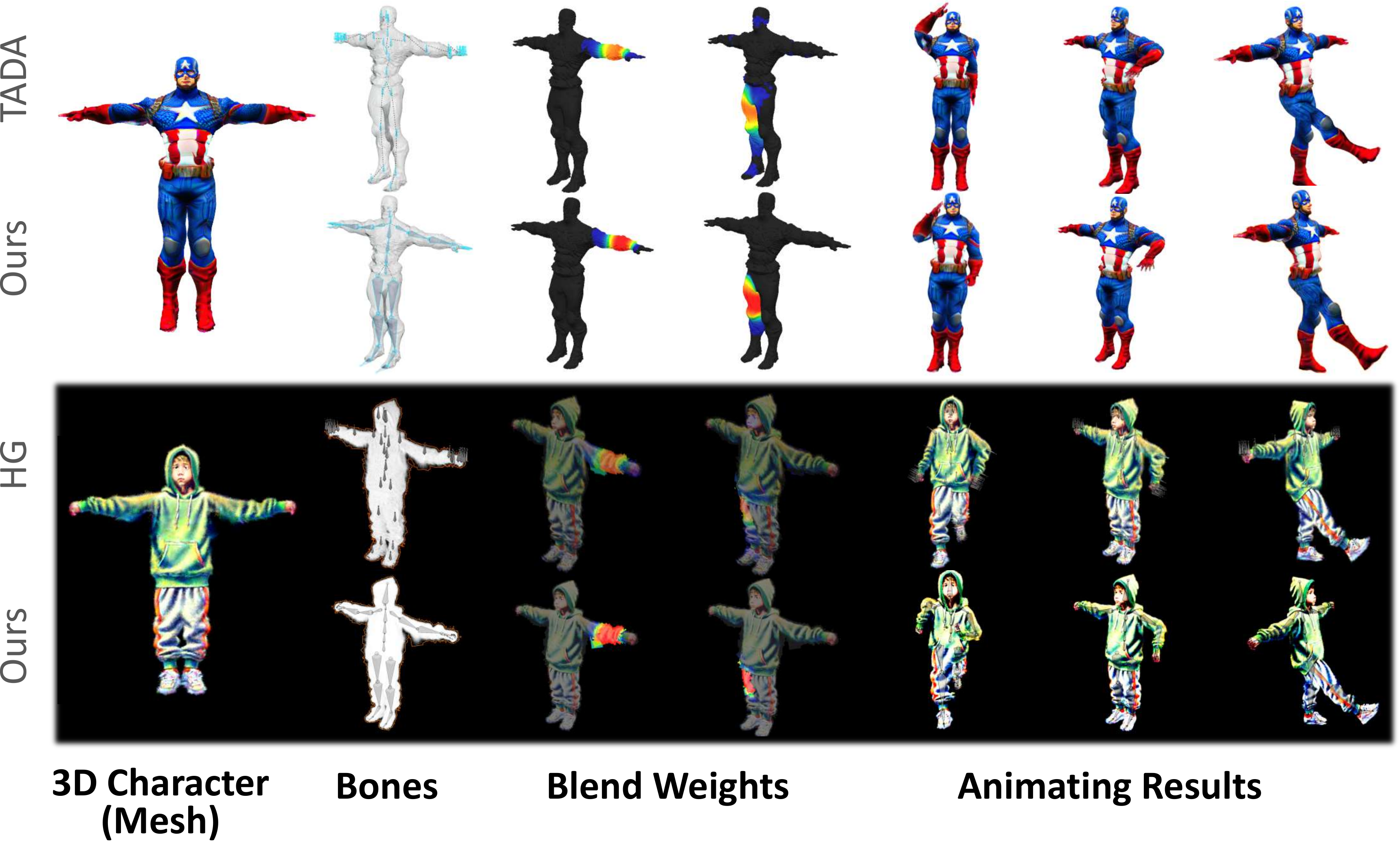}
      \vspace{-6px}
      \caption{
      \textbf{Comparison with TADA~\cite{liao2024tada} and HumanGaussian~\cite{liu2024humangaussian} (HG).}
      We use the generated meshes from TADA and 3D Gaussians from HG for comparison.
      Note that the skeletons of these two baselines are identical to the shape-specific SMPL~\cite{loper2015smpl} templates (without bone tail), with their weights interpolated from the template meshes.
      Zoom in to better view the details.
      }
      \label{fig:tada_hg}
\end{figure}

\begin{figure*}[t]
      \centering
      \includegraphics[width=1.0\linewidth]{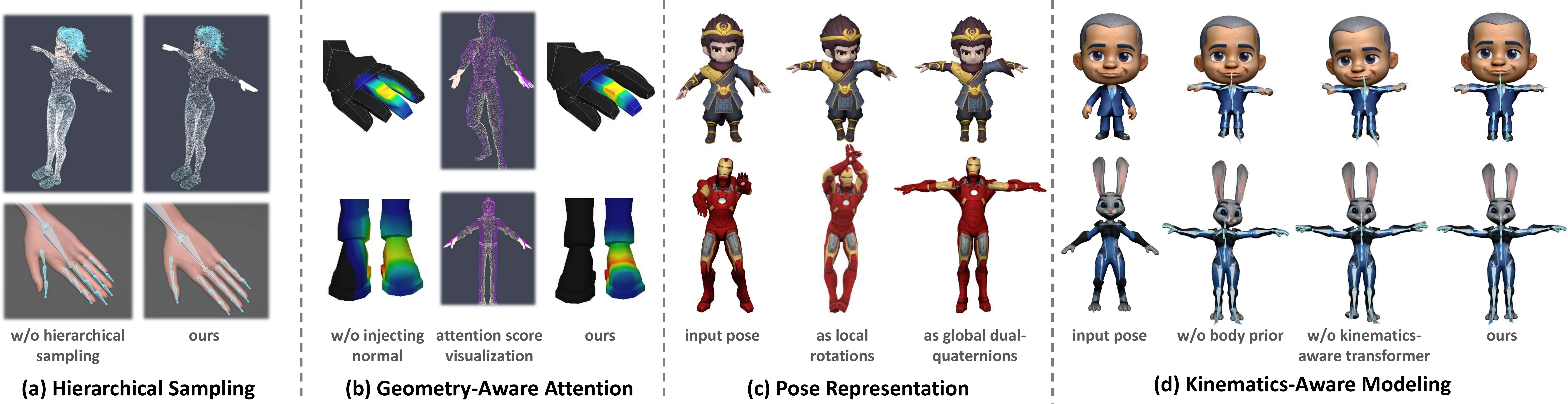}
      \vspace{-6pt}
      \caption{
      \textbf{Visualizations of some ablative experiments.}
      We show the effectiveness of the proposed modules and design choices by visualizing the predicted bones, blend weights, and pose transformations.
      }
      \label{fig:ablation}
      \vspace{-10pt}
\end{figure*}

In \cref{fig:meshy_tripo}, we compare the proposed method with two commercial software, Meshy~\cite{meshy} and Tripo~\cite{tripo}.
Meshy tends to enlarge the influence regions of skinning weights, leading to incorrect deformations of limbs. It also fails to extract the rest pose of some complex input shapes and animates the characters in a weird way.
Tripo does better in blend weights, but often produces inaccurate and unmatched bones. Plus, it also has difficulty determining the correct rest state for unusual input poses.
Moreover, both of the two cannot predict finger bones, resulting in coarse hand poses in the running scenario.
In contrast, our method produces complete and accurate bones as well as suitable blend weights, so the animations are also natural and fluid.

In \cref{fig:rignet}, we evaluate the auto-rigging method RigNet~\cite{xu2020rignet} with given test meshes.
It can be observed that RigNet often produces unnecessary topological structures, which is also manifested in its bone prediction errors from \cref{tab:rignet}.
Besides, its weights usually fail to maintain spatial smoothness, resulting in fragmented animating results.

In \cref{fig:tada_hg}, we further compare two generative works that perform text-to-3D-avatar generation. TADA~\cite{liao2024tada} produces textured meshes and HumanGaussian~\cite{liu2024humangaussian} chooses to use Gaussian splats. Our framework is directly compatible with both their representations. Since these two works generate shapes based on the well-defined SMPL~\cite{loper2015smpl} template, their joints' locations and weights are good enough for animating. Nevertheless, our method can achieve comparable, if not superior, quality using a template-free pipeline, without any prior knowledge of the character's shape or pose parameters. Furthermore, although these methods exhibit some deviations from the template mesh, they remain constrained by the preset body ratio of SMPL and are limited to shapes resembling realistic humans. In contrast, our method can be applied to a much wider range of character shapes.

\subsection{Ablation Study}
\label{sec:exp_ablation}

We conduct ablative experiments to validate the effectiveness of the proposed components and strategies, as shown in \cref{tab:ablation} and \cref{fig:ablation}.
It can be observed that the canonical transformation is vital for the proper convergence of our model, as it greatly reduces the distribution dispersion of inputs.
The large improvement brought by hierarchical sampling also indicates the importance of the proposed coarse-to-fine shape representation. As illustrated in \cref{fig:ablation} (a), sampling more in the fine-grained hand regions leads to a more accurate estimation of finger bones.
Besides, the injection of additional normal information via geometry-aware attention boosts the performance, especially in weight prediction. We visualize the attention score of the input point clouds in \cref{fig:ablation} (b), where brighter color indicates more attention to normals rather than coordinates. The model adaptively learns to rely more on normals in regions like inner thigh and between fingers since coordinates become less discriminative there.
Furthermore, accurately modeling pose transformations presents a complex challenge. When poses are predicted as local rotations rather than global dual quaternions, the model is limited to handling simpler inputs. Without a body prior loss, predicted limb and spine bones often exhibit slight but critical offsets, disrupting connectivity and symmetry. 
Additionally, in the absence of a structure-aware transformer, deeper-level bones, such as those in the fingers and feet, are susceptible to inaccuracies in pose-to-rest transformations. Their poses are heavily influenced by ancestral bones, where even minor inconsistencies can result in significant deformities.

\begin{table}[t]
  \centering
  \resizebox{1.0\linewidth}{!}{
    \begin{tabular}{l|cccccc}
    \toprule
     & IoU~$\uparrow$ & Precision~$\uparrow$ & Recall~$\uparrow$ & CD-J2J~$\downarrow$ & CD-J2B~$\downarrow$ & CD-B2B~$\downarrow$ \\
    \midrule
    RigNet~\cite{xu2020rignet} & 53.50\% & 47.27\% & 89.30\% & 6.63\% & 4.97\% & 2.88\% \\
    Ours & \textbf{82.50\%} & \textbf{81.07\%} & \textbf{90.31\%} & \textbf{4.49\%} & \textbf{3.32\%} & \textbf{1.58\%} \\
    \bottomrule
    \end{tabular}
  }
\vspace{-6pt}
\caption{\textbf{Quantitative comparison of skeleton prediction on the bipedal humanoid subset of the test dataset~\cite{xu2019predicting}.}
}
\label{tab:rignet}
\vspace{-5pt}
\end{table}

\begin{table}[t]
  \centering
  \resizebox{1.0\linewidth}{!}{
    \begin{tabular}{l|ccc}
    \toprule
     & Weights Error~$\downarrow$ & Joints Error~$\downarrow$ & Poses Error~$\downarrow$ \\
    \midrule
    w/o canonical transformation & 6.27\% & 9.80\% & 41.8\% \\
    w/o hierarchical sampling & 5.55\% & 2.42\% & 18.0\% \\
    w/o geometry-aware attention & 5.16\% & 2.20\% & 14.2\% \\
    w/o structure-aware transformer & - & 2.13\% & 14.9\% \\
    w/o body prior loss & - & 2.13\% & 14.0\% \\
    w/o global pose representation & - & - & 35.3\% \\
    \textbf{Ours} & \textbf{4.70\%} & \textbf{2.11\%} & \textbf{13.6\%} \\
    \bottomrule
    \end{tabular}
  }

\vspace{-6pt}
\caption{\textbf{Ablation studies on the test split of the Mixamo dataset.}
We report the percentage error of animation assets.
}
\label{tab:ablation}
\vspace{-8pt}
\end{table}

\section{Conclusion and Discussion}
\label{sec:conclusion}
In this paper, we propose a novel framework for animation-ready 3D character production.
To address non-trivial challenges and practical limitations of existing methods, we develop several elaborate modules, including the coarse-to-fine shape representation, the particle-based shape autoencoder, and the structure-aware modeling of bones.
Putting all these together, we provide an out-of-the-box and efficient solution to animating any 3D character.
Comprehensive experiments demonstrate the superiority of our method and its considerable potential for future investigation.

Despite the merits, there is still room for improvement.
Making the empirical hierarchical sampling more adaptive may be a meaningful future direction.
It is also a promising avenue to extend the proposed framework to non-bipedal characters by proposing a more flexible way of modeling the topological structures of bones.

{
    \small
    \bibliographystyle{ieeenat_fullname}
    \bibliography{main}
}

\clearpage
\setcounter{page}{1}
\maketitlesupplementary

\makeatletter
\renewcommand{\theHsection}{arabicsection.\thesection}
\renewcommand \thesection{S\@arabic\c@section}
\setcounter{section}{0}
\renewcommand \thetable{S\@arabic\c@table}
\setcounter{table}{0}
\renewcommand \thefigure{S\@arabic\c@figure}
\setcounter{figure}{0}
\renewcommand \theequation{S\@arabic\c@equation}
\setcounter{equation}{0}
\makeatother

\section{Formulation of Low-Rank Dynamics}

As discussed in \cref{sec:pre}, modeling the dynamics of an object typically requires a per-timestamp deformation of all the particles (\ie, vertices for meshes, points for point clouds, and splats for 3D Gaussians) that make up its geometry.
Suppose that we have an object composed of $N$ particles and a desired dynamic sequence of $T$ timestamps. The temporal deformations of all particles can be represented by a matrix $\bm{D} \in \mathbb{R}^{T\times N \times d}$, where $d$ is the degrees of freedom (\eg, $d=6$ for rigid transformations).
Generally, $\bm{D}$ has a lot of redundancy when representing real-world dynamics. Therefore, we would like to seek a low-rank approximation of it. 
Mathematically, the low-rank decomposition of a matrix like $\bm{D}$ has two different forms expressed by
\begin{align}
    \bm{D}^{T\times N \times d} \approx \bm{B}_t^{T \times K \times d} \bm{W}_s^{K \times N},
    \label{eq:lowrank}
    \\
    \bm{D}^{T\times N \times d} \approx \bm{B}_s^{K \times N \times d} \bm{W}_t^{T \times K},
    \label{eq:lowrank2}
\end{align}
where $K$ is the desired rank.
We call the matrix $\bm{B}$ with dimension $d$ as \textit{basis}, and the other one $\bm{W}$ as \textit{weight}.
\cref{eq:lowrank,eq:lowrank2} both decouple the temporal and spatial dimensions by splitting them into basis and weight. We then name the ones with temporal dimension ($T$) as \textit{temporal basis} $\bm{B}_t$ and \textit{temporal weight} $\bm{W}_t$. Correspondingly, the matrices with spatial dimension ($N$) are named as \textit{spatial basis} $\bm{B}_s$ or \textit{spatial weight} $\bm{W}_s$.

In fact, both decomposition forms have been playing important roles in the applications of dynamic modeling.
For example, \cref{eq:lowrank} is applied in the sparse control paradigms~\cite{huang2024scgs,lei2024mosca,wang2024som} and the linear blend skinning (LBS)~\cite{loper2015smpl,pavlakos2019smplx} algorithm, where $\bm{B}_t$ is interpreted as the transformations/poses of control nodes/body joints, and $\bm{W}_s$ is the blend weights of the nodes or joints.
When applying \cref{eq:lowrank2}, $\bm{B}_s$ and $\bm{W}_t$ are interpreted as the blend shapes and their corresponding weights for the per-timestamp linear combination~\cite{li2017flame,li2021learning,zielonka2024gem,das2024npgs}.

In this work, we focus on the low-rank form of \cref{eq:lowrank} in modeling dynamic characters. The reason is twofold.
First, the temporal weight $\bm{W}_t$ in \cref{eq:lowrank2} typically requires as much motion data as possible for one object to find a capacious low-rank space, which is prohibitively difficult in practice. In contrast, the spatial weight $\bm{W}_s$ can be supervised much more easily by the well-defined blend weights in existing 3D models.
Second, considering the availability of rich motion resources~\cite{mixamo,lin2023motionx} and powerful motion generation methods~\cite{jiang2024motiongpt,chen2023executing}, there is no need to model the temporal basis $\bm{B}_t$ from scratch. What we have to do is finding a transformation from the input character pose to a pre-defined rest pose, and then any desired animation is within easy reach.

Note that theoretically, joint/bone positions are only proxies or interfaces for the low-rank terms and are not indispensable to the dynamic modeling, as expressed by \cref{eq:lowrank}. Consequently, some related works~\cite{wang2024som} choose to model such proxies in an implicit way.
However, we still include the bones as one of the desired animation assets in our work for a self-contained and artist-friendly representation compatible with existing animating pipelines. Furthermore, the explicit existence of bones can bring much convenience when applying body priors to assist the optimization, as introduced in \cref{sec:kinematics}.

\section{Implementation Details}

\begin{figure*}[t]
      \centering
      \includegraphics[width=\linewidth]{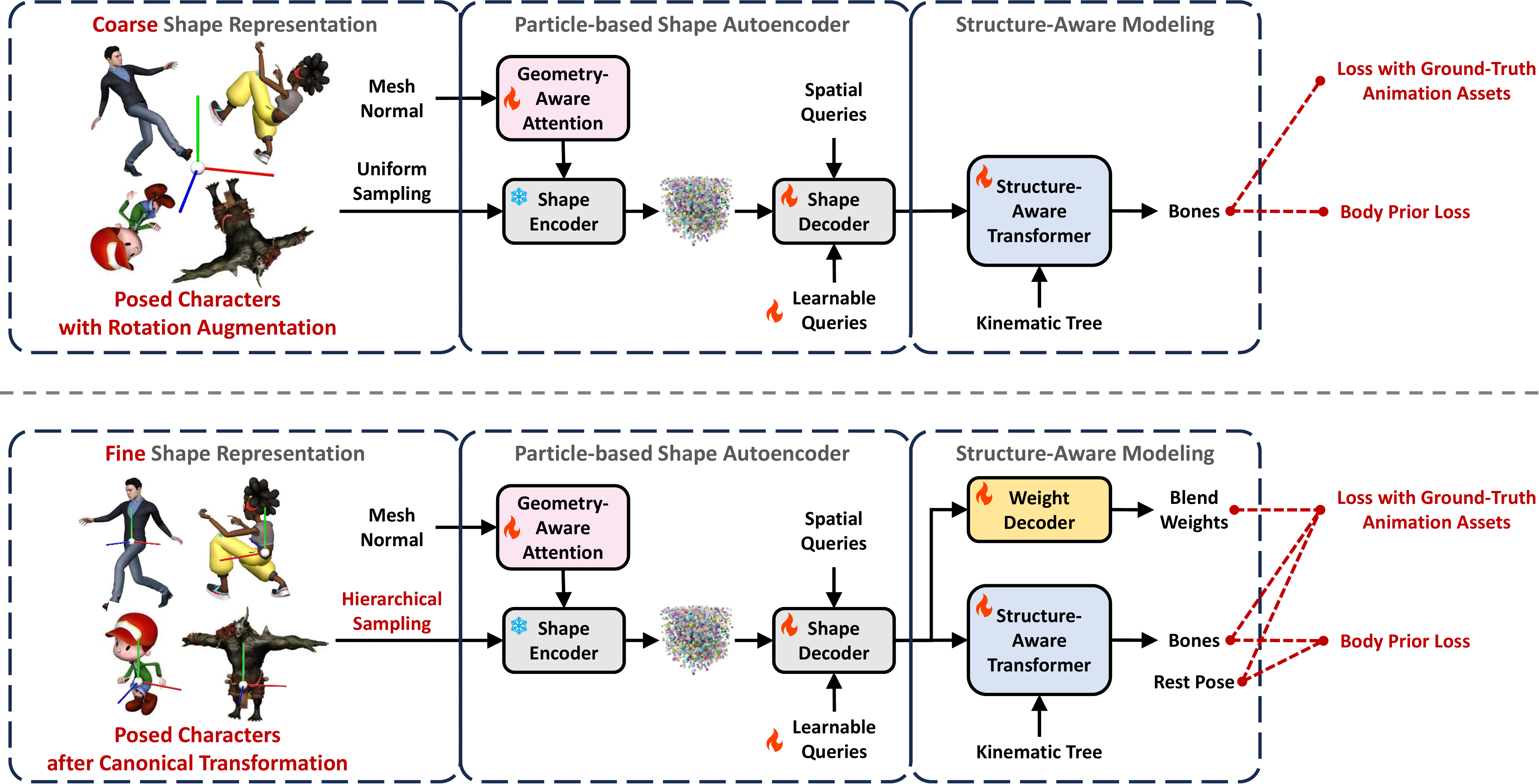}
      \caption{
      \textbf{The coarse (upper) and fine (lower) stages of training our framework.}
      In the coarse stage, the input shape is uniformly sampled and only the bone positions are predicted. We apply data augmentation to the inputs via random 3D rotations, so that the coarse model is generalizable to global transformations of in-the-wild cases with an acceptable accuracy.
      In the fine stage, we apply canonical transformation and hierarchical sampling to the shapes in advance based on the ground-truth bone positions.
      Then during inference, a 3D character is fed into the coarse framework to get its bone positions, which guide the establishment of coarse-to-fine shape representation later in the fine framework.
      Note that the body prior loss (\cref{sec:kinematics}) is directly applied to the bone positions. As for pose prediction, we take the ground-truth bones as a proxy and use the predicted pose to transform them, thereby indirectly affecting the pose optimization.
      }
      \label{fig:pipeline_coarse_fine}
\end{figure*}

\subsection{Coarse-to-Fine Shape Representation}

To address the limitations of the proposed framework and boost the performance, we introduce an additional design at the input side of the autoencoder, \ie, the coarse-to-fine shape representation (\cref{sec:sampling}).
In addition to the overall framework presented in \cref{fig:pipeline}, we include two separate pipelines in \cref{fig:pipeline_coarse_fine} here for a clearer illustration of the coarse and fine training stages.
In the following content, we will provide more details of this part, particularly regarding the motivations and implementation choices.

With the particle-based shape autoencoder, our method can already produce fairly good results of blend weights. However, the joint outputs are still unsatisfying in fine-grained regions like the hands. Meanwhile, the pose prediction can hardly converge.
We attribute these issues to the ambiguity of the input points and treat the lite training process (taking uniformly sampled points as input and only predicting bone positions, as shown in \cref{fig:pipeline_coarse_fine} upper) as a coarse stage that provides rough but valuable localization information about the input character. 
To be specific, we exploit some of the coarse joint locations to gain a finer shape representation by applying two different strategies, \ie, the canonical transformation and the hierarchical sampling.

\parsection{Canonical transformation}
Although we have normalized the input shapes to align their scales, they still differ in global transformations. Since the poses we want to predict are relative to the fixed origin, the same pose can have huge numerical differences under different coordinate systems. This will lead to a dramatically increased difficulty in pose prediction.
Therefore, we move and rotate the entire shape to a canonical position and orientation before sampling.
In practice, we choose this transformation based on an empirically selected datum plane (referred to as the \textit{hip plane}) determined by three joint positions, \ie, the hip (root of the kinematic tree) and two upper thighs, which can be accurately located even in the coarse stage.
The canonical transformation is applied so that:
1) the hip is located at the origin;
2) the normal of the hip plane is aligned with the z-axis;
3) the vector from the right to the left thigh is parallel to the x-axis.
This process simplifies the input shape's spatial distribution and eliminates the pose representation's ambiguity.
Furthermore, the chosen canonical transformation ensures a consistent upright and front-facing orientation of the input character, which is a common prerequisite of many auto-rigging methods~\cite{xu2020rignet,ma2023tarig,li2021learning}. By integrating the coarse localization, we now automate this preprocessing step, enabling our framework to effectively handle the inputs regardless of their initial spatial configuration (positions, rotations, and scales).

\parsection{Hierarchical sampling}
Some parts of the input character, \eg, the hands, present fine-grained details within small regions, which typically demand additional resources for accurate processing. However, the uniform sampling applied to the input shape, as well as the subsequent farthest point sampling (FPS) algorithm, usually results in sparsely distributed sample points on hands, which are far from enough to describe the geometry of fingers.
Theoretically, increasing the number of sampling points $N$ and the downsampling number $M$ can bring a larger representation capacity for the entire geometry including the hand regions, but that will significantly add to the computational overhead.
For efficiency, we choose to keep the total sampling number unchanged and instead leverage the coarse prediction of hand joints.
Specifically, we replace the uniform sampling with a hierarchical approach that ensures a designated proportion of sample points are distributed on both hands.
Since the FPS algorithm in the shape encoder disrupts the non-uniformity of samples, we adapt it to a hierarchical algorithm as well.

\subsection{Networks and Hyperparameters}

For the shape autoencoder, we use a point cloud of size $N=32768$ as the input. All the points are sampled on the surface of the input mesh. In the fine training or the inference stage equipped with the hierarchical sampling, 50\% of the points are uniformly sampled and the other 50\% are sampled near the hand joints.
The hyperparameters of the shape autoencoder are consistent with the original setting of 3DShape2VecSet~\cite{zhang20233dshape2vecset}, which internally uses shape latents $\bm{F} \in \mathbb{R}^{M \times C}$ with $M=512$ and $C=512$ for the neural field.
Note that in \cite{zhang20233dshape2vecset}, $\bm{F}$ can either be learnable embeddings or initialized by applying farthest point sampling (FPS) on the input point cloud. We choose the latter implementation in this work.
Our framework is trained on 8 NVIDIA A100 GPUs.
The learning rate is linearly increased to $1e-4$ within the first 1\% iterations (warm-up), and then gradually decreased using the cosine decay schedule until reaching the minimum value of $1e-5$.

\subsection{Body Prior Losses}
The implementation of the body prior losses (introduced in \cref{sec:kinematics}) involves:
1) defining a prior-based bone pair list,
and
2) applying $L_1$-based loss to their predicted positions or directions.
For example, to encourage bone connectivity, we define a list of bone pairs that should be connected based on the predefined topology, select their predicted head-tail positions, and penalize their pair-wise $L_1$ distances.

\section{Data Details}

\subsection{Mixamo Dataset}

\begin{figure}[t]
      \centering
      \includegraphics[width=\linewidth]{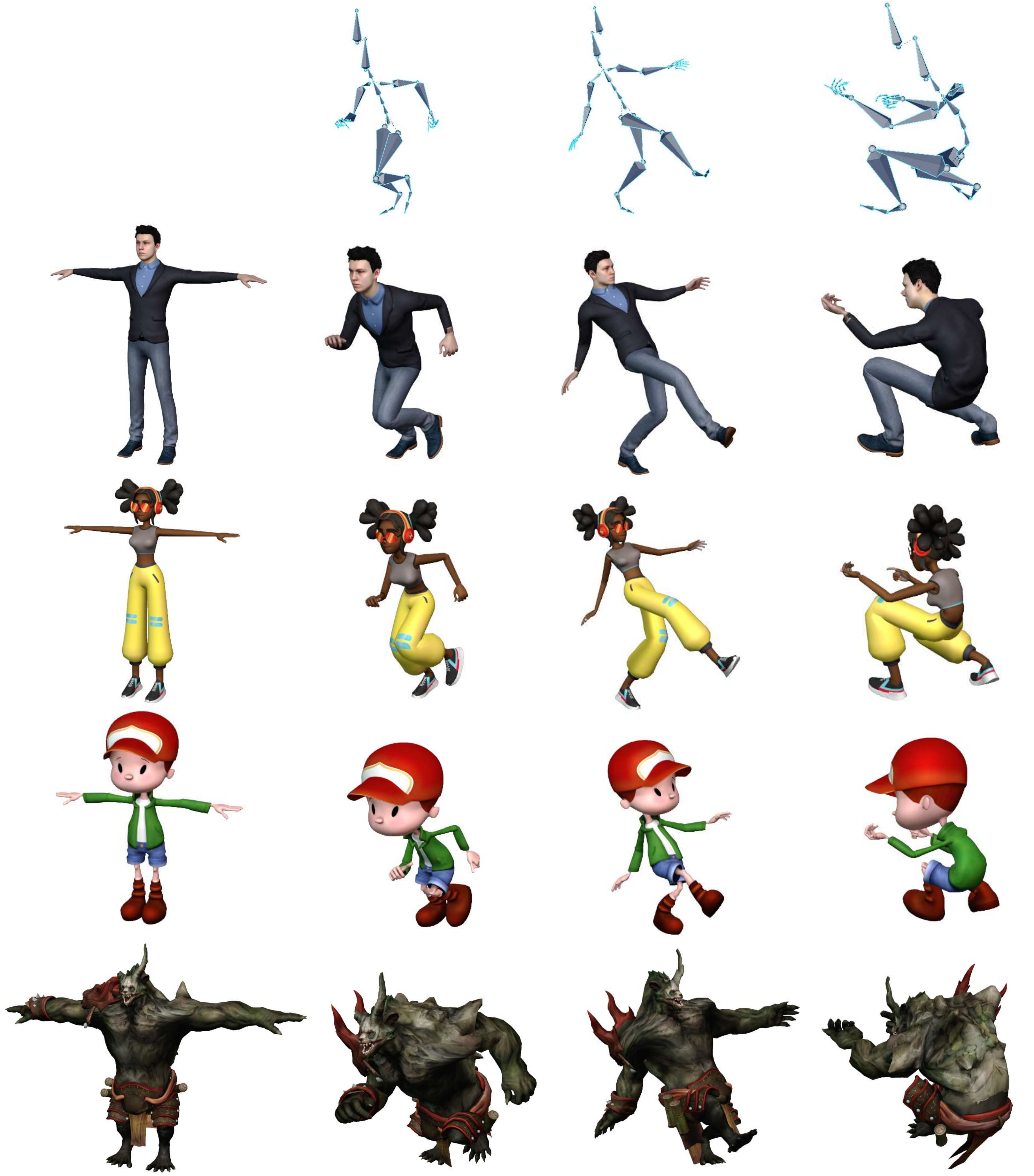}
      \caption{
      \textbf{Some samples from the collected Mixamo~\cite{mixamo} dataset.}
      The dataset contains bipedal humanoids with different shapes, ranging from realistic humans to cartoon or fantasy creatures. Each character is preprocessed to be animatable by any of the motion sequences.
      The proposed framework is trained on this dataset.
      }
      \label{fig:mixamo_dataset}
\end{figure}

To obtain sufficient training data of 3D characters with high-quality geometry and animation assets, we collect the artist-designed 3D models from Mixamo~\cite{mixamo} to form a dataset, which comprises texture meshes of 95 exquisite characters (each has an average of 15000 vertices), along with 2453 diverse motion sequences (each has an average of 200 frames).
All the characters are preprocessed to share a standard skeleton structure with $K=52$ bones (the leaf bones are removed).
For characters with non-standard skeletal structures originally, the blend weights of those bones are transferred to their topologically nearest ancestral bones within the standard skeleton.
If some standard bones are missing in a character (\eg, armless person), we mask their corresponding values (weight channels and bone-wise attributes) in loss computing.
We use 95\% of the data for training and the remaining 5\% for for validation.
During each training iteration, we randomly choose a character-motion pair to get the corresponding animation assets, resulting in an effective dataset size equivalent to over 40 million frames.
In \cref{fig:mixamo_dataset}, we show some example characters and motions from the Mixamo dataset.

\subsection{VRoid Dataset}

\begin{figure}[t]
      \centering
      \includegraphics[width=\linewidth]{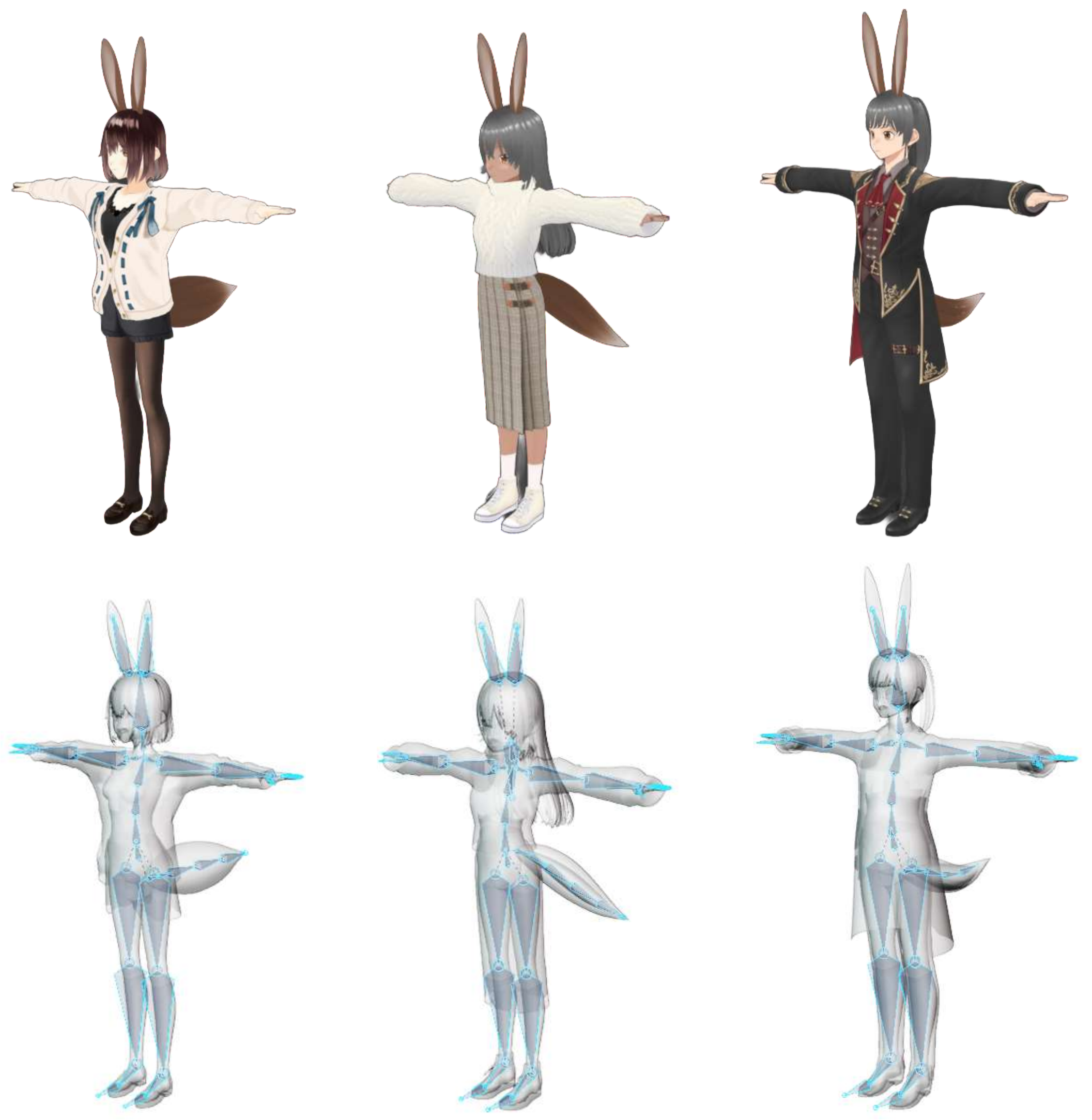}
      \caption{
      \textbf{Some samples of the anime characters with additional accessories for the extra-bone fine-tuning.}
      These characters are all manually created using VRoid Studio~\cite{vroid} and then preprocessed to be compatible with the standard skeleton definition of Mixamo~\cite{mixamo}.
      }
      \label{fig:vroid_dataset}
\end{figure}

\begin{figure*}[t]
      \centering
      \includegraphics[width=\linewidth]{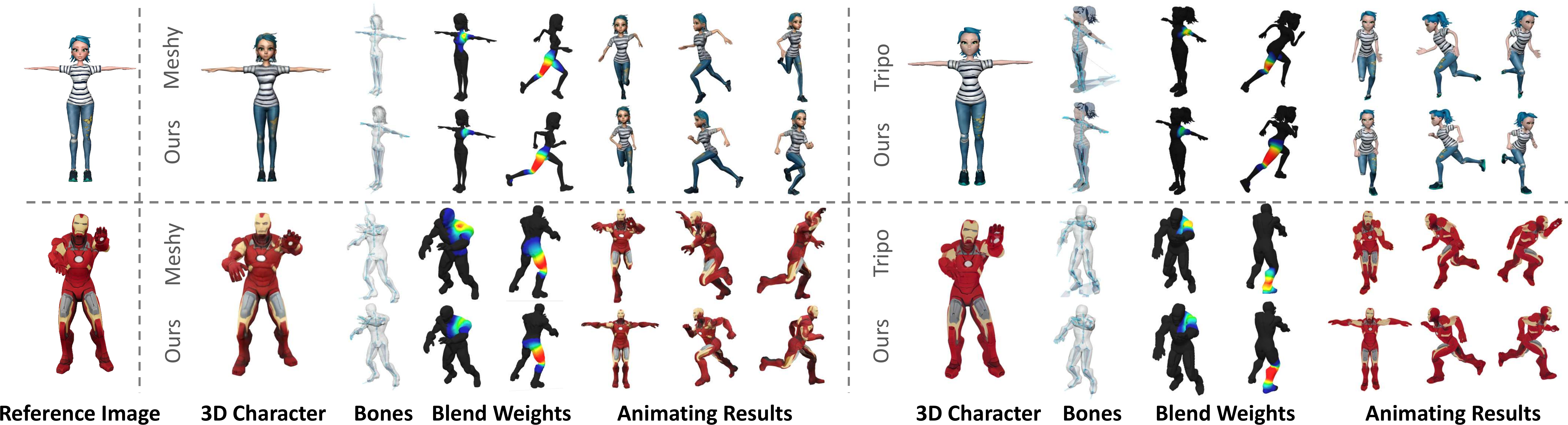}
      \caption{
      \textbf{Additional Comparison with generative 3D methods, \ie, Meshy~\cite{meshy} and Tripo~\cite{tripo}.}
      We feed them the same image as reference and compare the performance based on their generated 3D models respectively.
      The blend weights of two joints, \ie, Left Shoulder and Right Leg, are visualized.
      Given that these baselines can only apply preset motions and their rest-pose models cannot be exported, we apply a similar ``running'' sequence to all the methods for fair comparison.
      For non-rest cases, the T-pose models predicted by our method are included as the front-view animating results.
      Zoom in to better view the details.
      }
      \label{fig:meshy_tripo_more}
\end{figure*}

\begin{figure*}[t]
      \centering
      \includegraphics[width=\linewidth]{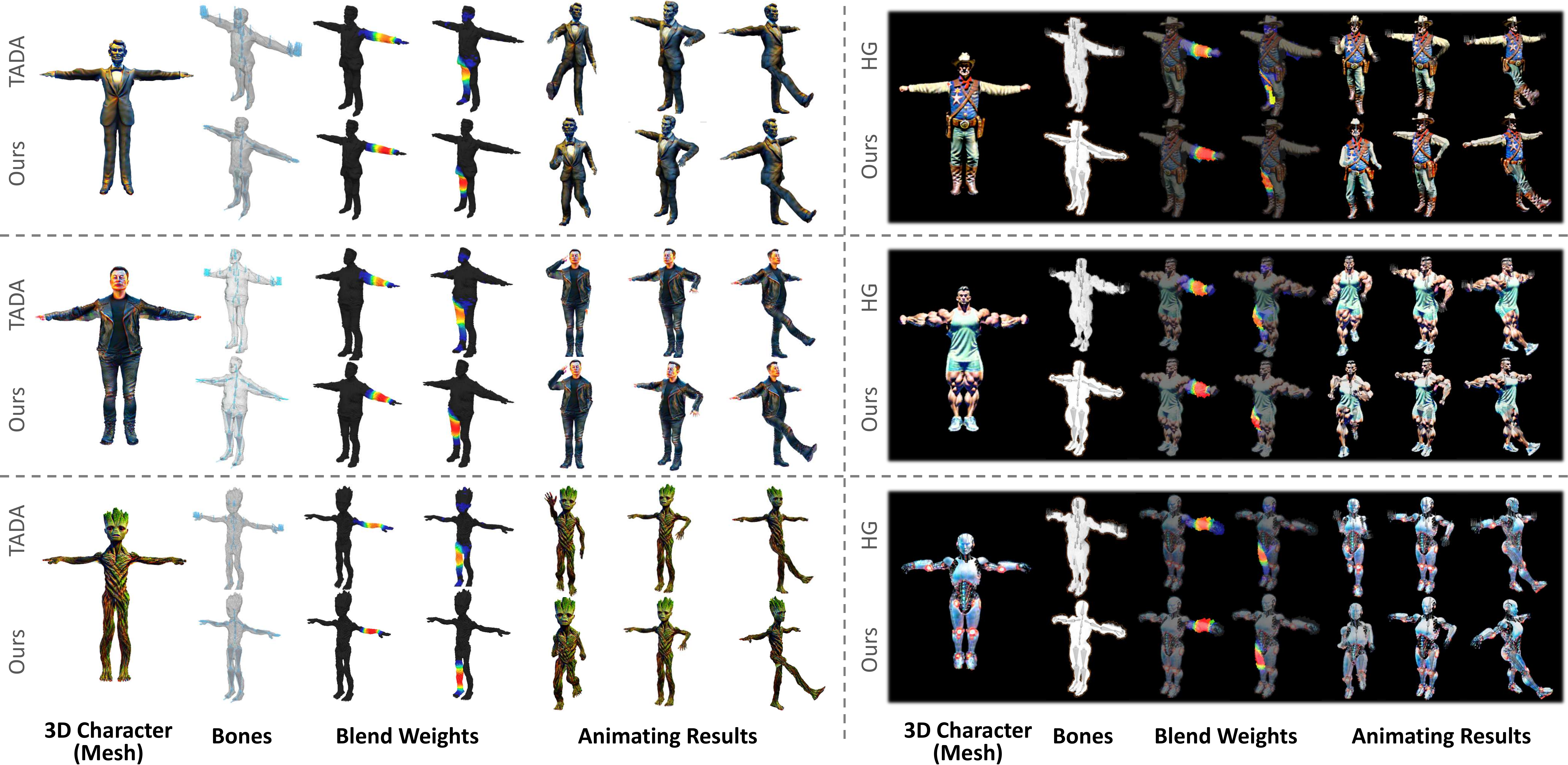}
      \caption{
      \textbf{Additional comparison with template-based avatar generation methods, \ie, TADA~\cite{liao2024tada} and HumanGaussian~\cite{liu2024humangaussian} (HG).}
      We use the generated meshes from TADA and 3D Gaussians from HG for comparison.
      Note that the skeletons of these two baselines are identical to the shape-specific SMPL~\cite{loper2015smpl} templates (without bone tail), with their weights interpolated from the template meshes.
      Zoom in to better view the details.
      }
      \label{fig:tada_hg_more}
\end{figure*}

\begin{figure*}[t]
      \centering
      \includegraphics[width=\linewidth]{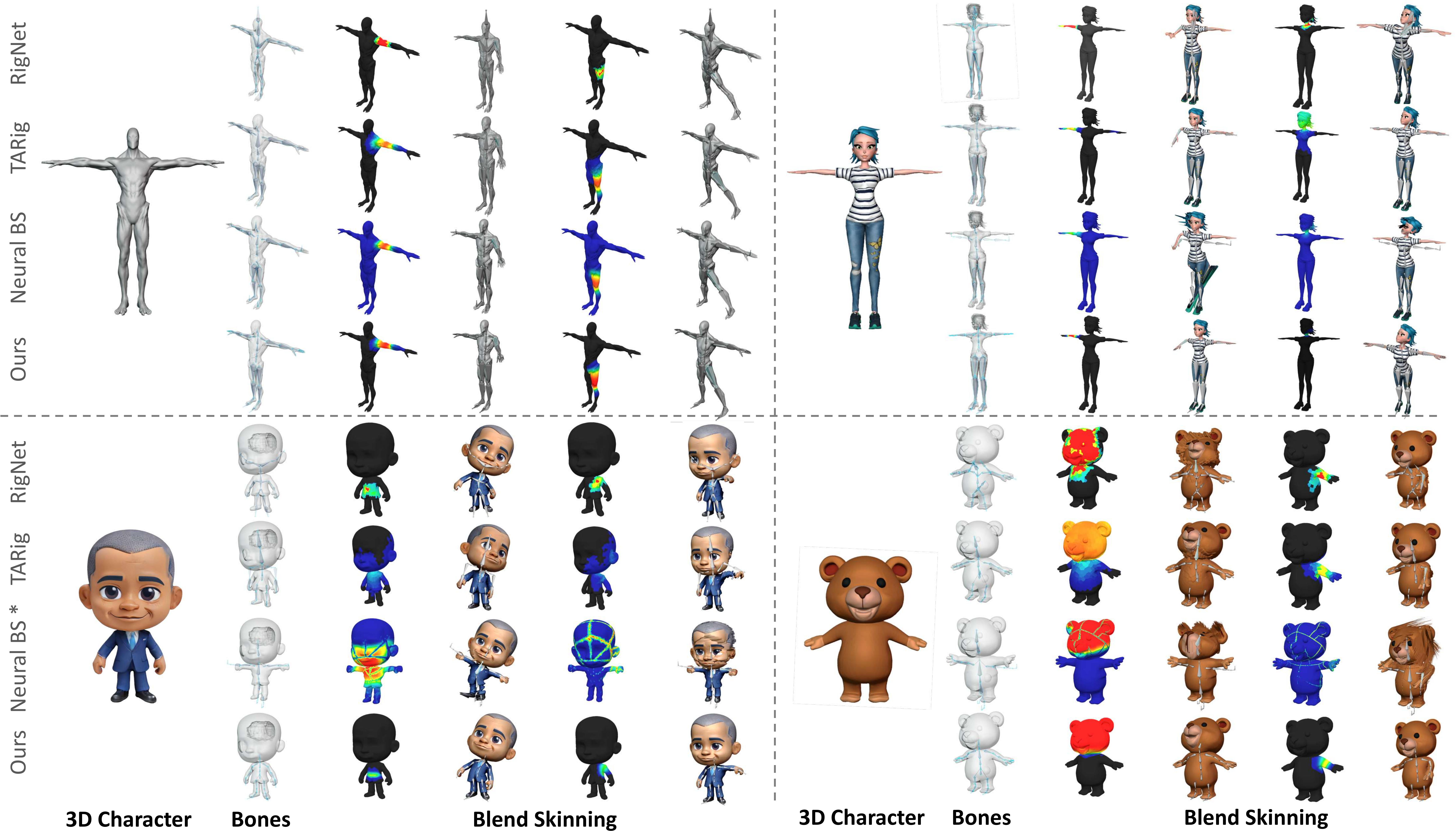}
      \caption{
      \textbf{Additional comparison with auto-rigging methods, \ie, RigNet~\cite{xu2020rignet}, TARig~\cite{ma2023tarig}, and Neural Blend Shapes~\cite{li2021learning} (Neural BS).}
      We visualize the blend weights of selected joints and manually deform them to assess the impact of rigging quality on skinning results.
      *: Neural Blend Shapes only support T-pose inputs, so for the non-rest cases (lower two), we feed it the T-pose meshes transformed by our pose-to-rest predictions.
      }
      \label{fig:rignet_more}
\end{figure*}

\begin{figure*}[t]
      \centering
      \includegraphics[width=0.9\linewidth]{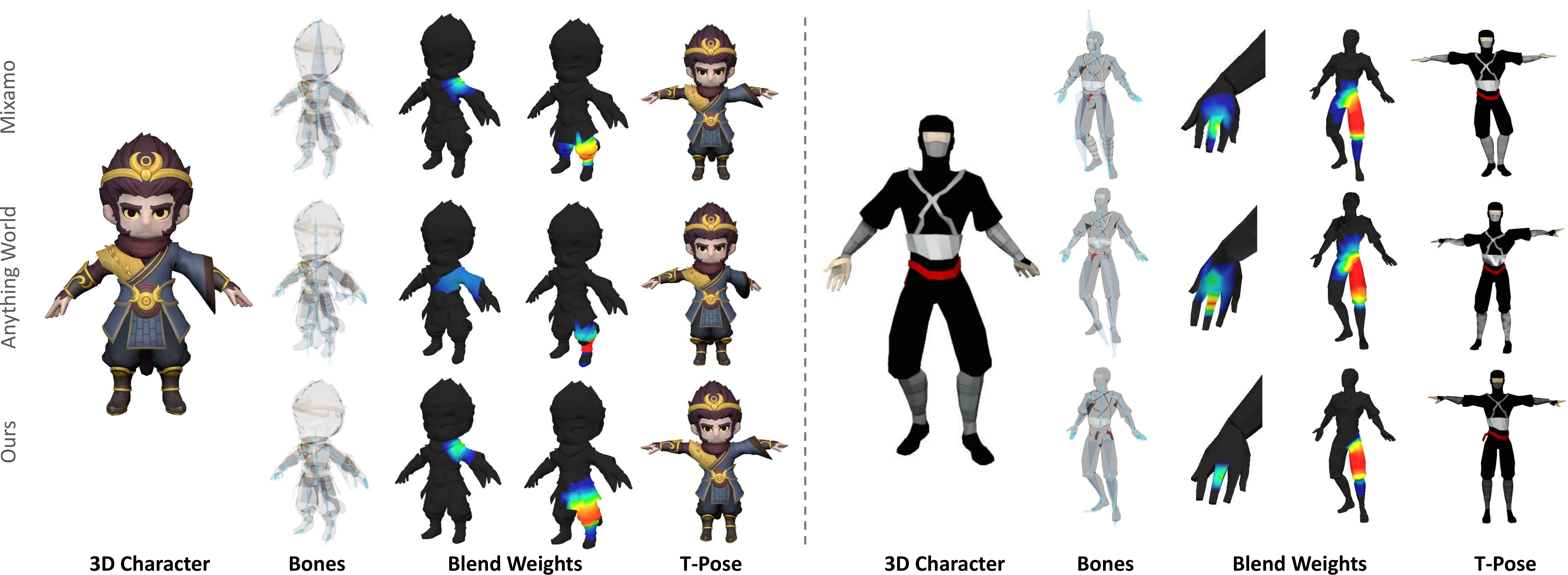}
      \caption{
      \textbf{Comparison with commercial auto-rigging software, \ie,  Mixamo~\cite{mixamo} and Anything World~\cite{anythingworld}.}
      Note that these two tools can only deal with simple input poses (T- or A-pose is recommended) and often raise errors when faced with complex ones.
      }
      \label{fig:mixamo_aw}
\end{figure*}

\begin{figure*}[t]
      \centering
      \includegraphics[width=0.85\linewidth]{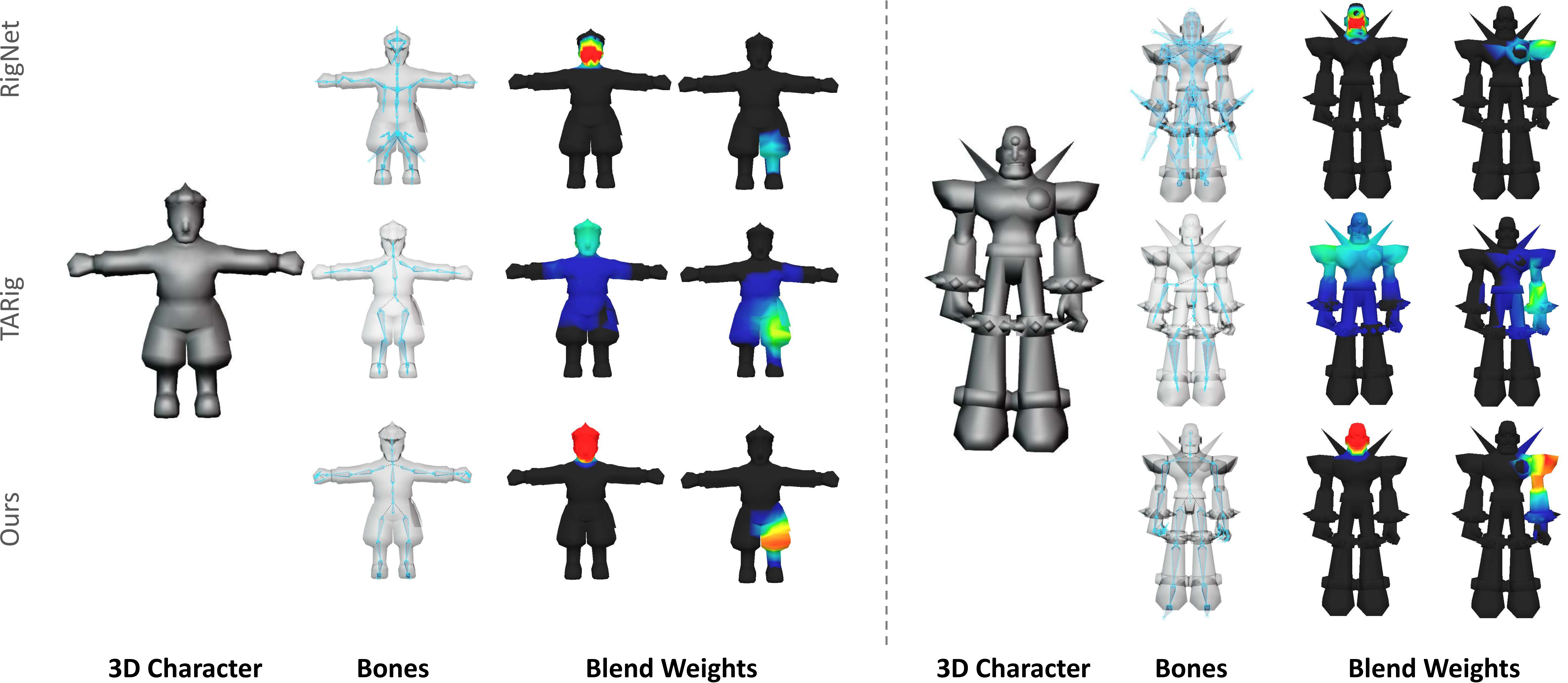}
      \caption{
      \textbf{Qualitative comparison with RigNet~\cite{xu2020rignet} and TARig~\cite{ma2023tarig} on cases from the test split of ``ModelsResource-RigNetv1'' dataset~\cite{xu2019predicting}.}
      While both baselines are exactly trained on this dataset and ours are not, we still achieve the best performance.
      }
      \label{fig:rignet_dataset}
\end{figure*}

Our training framework can also be extended to different skeleton topologies.
To demonstrate this capacity, we use VRoid Studio~\cite{vroid} to manually create 35 different anime characters (30 for training and 5 for validation) with additional accessories including two rabbit-like ears and a fox-like tail, as shown in \cref{fig:vroid_dataset}.
These characters differ in body shape, clothes, hair style, and the shape of ears and tails.
They are all preprocessed to share the same skeleton definition of Mixamo~\cite{mixamo} but with extra bones of the accessories. Therefore, they can also be animated with any of the Mixamo motion sequences during the extra-bone training (\ie, fine-tuning the final layer of the weight decoder and the extra learnable queries, based on the standard-skeleton model pretrained on the Mixamo dataset).
The experiments show that with our framework, 30 training characters are sufficient to obtain a good model that produces promising predictions for extra bones.

\section{More Results}

\subsection{Analysis of Efficiency}
\begin{figure}[t]
      \centering
      \includegraphics[width=\linewidth]{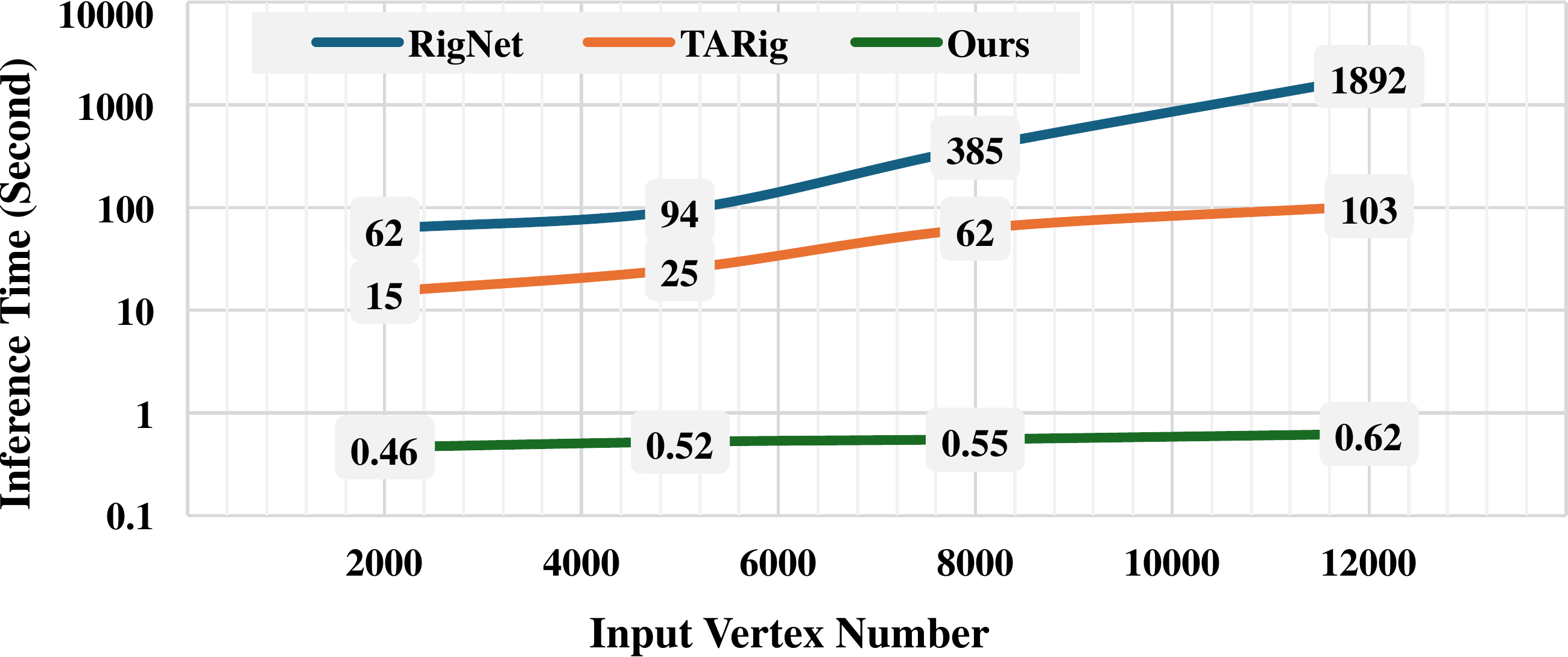}
      \caption{
      \textbf{Comparison of the inference time cost with increasing number of vertices as the input.
      }
      }
      \vspace{-3mm}
      \label{fig:eff}
\end{figure}

We compare the inference speed with inputs of different vertex numbers in \cref{fig:eff}.
Our method maintains sub-second time cost even with 10,000+ vertices, achieving 1000x and 100x speedups over RigNet~\cite{xu2020rignet} and TARig~\cite{ma2023tarig} respectively. While TARig accelerates RigNet's iterative process through feed-forward regression, we further introduce a particle-based shape encoder that bypasses mesh connectivity processing while achieving better performance. This architecture also enables point sub-sampling, making our approach particularly efficient for high-resolution inputs.

\subsection{Additional Comparison Results}

\parsection{More comparison cases}
In addition to \cref{fig:meshy_tripo,fig:tada_hg}, we exhibit more cases of comparison with the baselines, including Meshy~\cite{meshy} and Tripo~\cite{tripo} (\cref{fig:meshy_tripo_more}), TADA~\cite{liao2024tada} and HumanGaussian~\cite{liu2024humangaussian} (\cref{fig:tada_hg_more}).
The results can prove the effectiveness, robustness, and generalizability of the proposed framework.

\parsection{Comparison with more auto-rigging methods}
For comparison with existing auto-rigging methods, we include more cases and compare with two more baselines here in \cref{fig:rignet_more} (in addition to \cref{fig:rignet}), \ie, Neural Blend Shapes~\cite{li2021learning} and TARig~\cite{ma2023tarig}.
Neural Blend Shapes only supports T-pose inputs and has quite limited generalizability to shapes different from the SMPL mesh.
TArig's skeleton predictions are better than Neural Blend Shapes and RigNet~\cite{xu2020rignet}, but its blend weights still cannot meet the standard of practical application, producing spatial unsmoothness when deforming the meshes.
Besides, all three baselines are unable to produce fine-grained hand bones, while our method handles the fingers well.

Furthermore, the commercial software, Mixamo~\cite{mixamo} and Anything World~\cite{anythingworld}, rely much on the symmetry or pose simplicity (\eg, T-pose and A-pose) of the inputs and will raise errors when faced with complex ones.
Therefore, we compare them separately on some additional cases here in \cref{fig:mixamo_aw}.
It can be observed that
Anything World produces significant errors when extracting the bones of the left arm for the character with a tail (left case).
Meanwhile, Mixamo fails to fail to distinguish the left and right sides of the ninja (right case) and produces a mirrored skeleton.
Moreover, when faced with a non-rest input character like this ninja, the predicted pose-to-rest transformations of Mixamo and Anything World both suffer from unnatural deformations, while our results remain good.

\parsection{Qualitative comparison on the ModelsResource dataset~\cite{xu2019predicting}}
We also evaluate the proposed framework on the bipedal-humanoid subset of the ``ModelsResource-RigNetv1'' dataset~\cite{xu2019predicting}. \cref{fig:rignet_dataset} shows some cases for qualitative comparison with RigNet~\cite{xu2020rignet} and TARig~\cite{ma2023tarig}. Both of these two baselines are trained on the aforementioned dataset, but our model has never encountered a similar data distribution during training. Despite this, our method still achieves the best quality in rigging and skinning, demonstrating its strong generalizability.

\subsection{Additional Visualizations}

\begin{figure}[t]
      \centering
      \includegraphics[width=\linewidth]{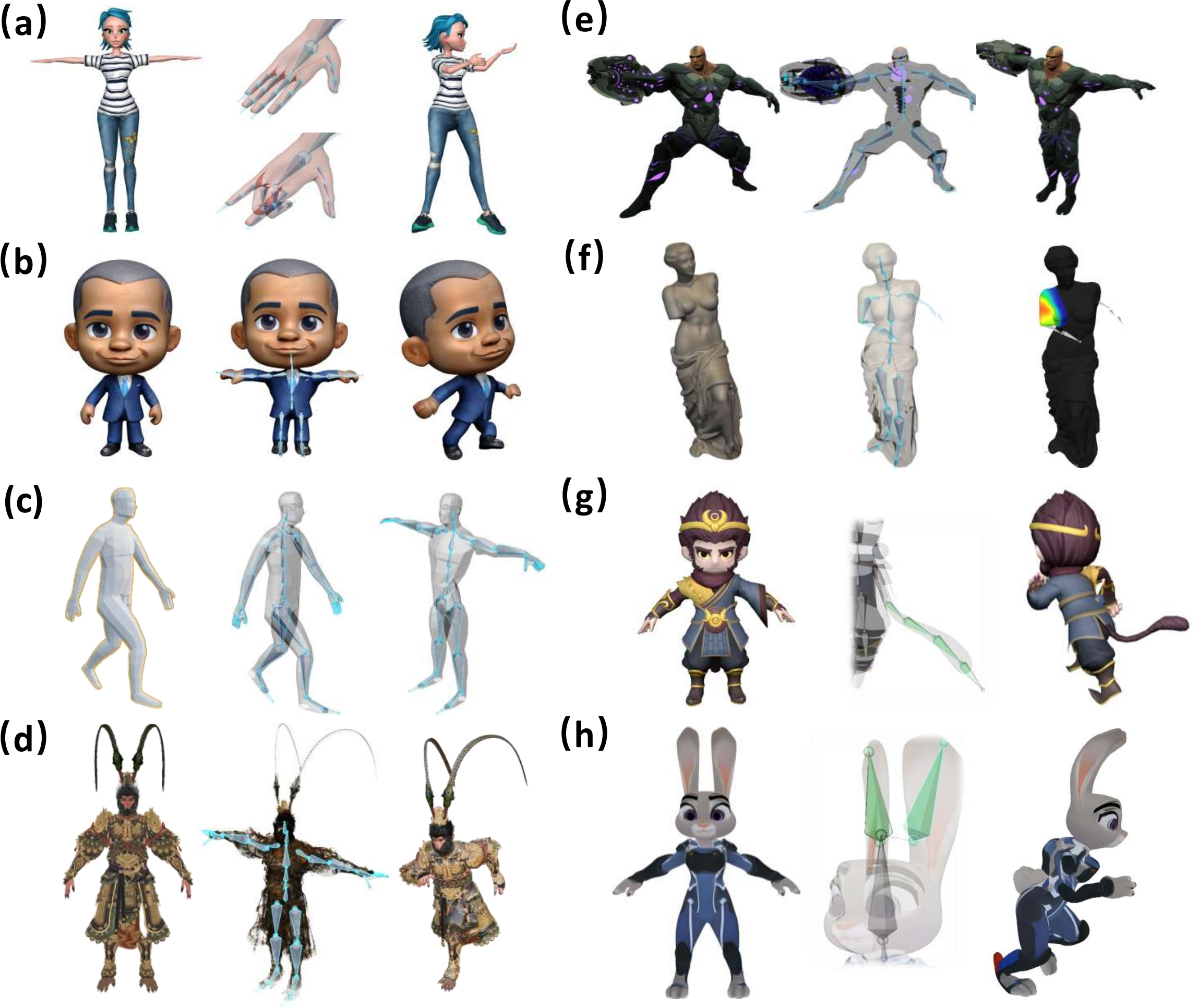}
      \caption{
      \textbf{Results of more tricky cases to demonstrate the advantage of our method.}
      (a) Fine-grained control of fingers; (b) Capacity of unusual shapes; (c) Complex input poses; (d) Efficiency for high polygon models; (e) Support of asymmetric inputs; (f) Adaptation to non-existing bones; (g) \& (h): Extension to extra bones (\eg, long ears and tails).
      }
      \label{fig:tricky}
\end{figure}

\begin{figure}[t]
      \centering
      \includegraphics[width=\linewidth]{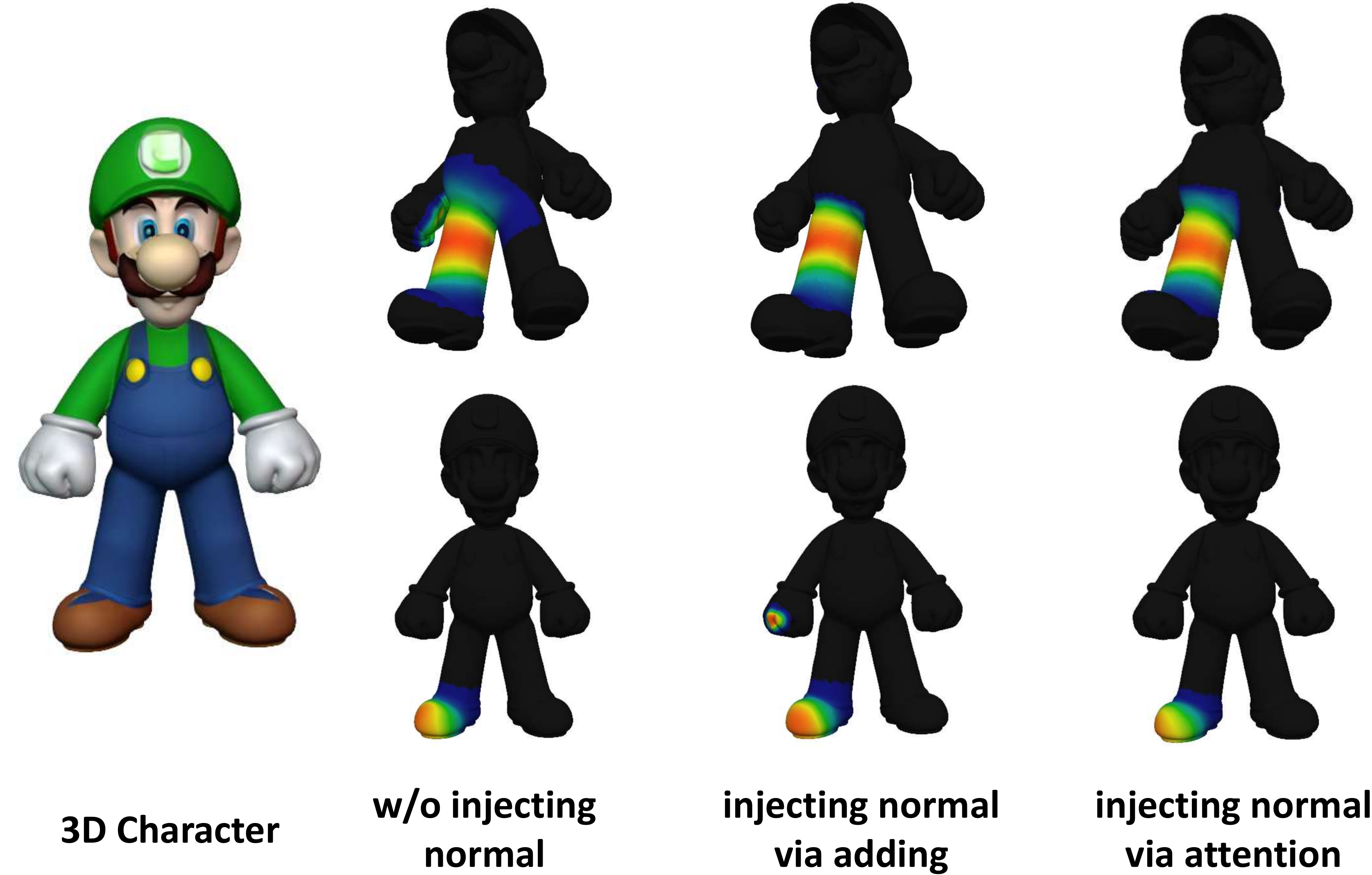}
      \caption{
      \textbf{Qualitative analysis of our geometry-aware attention module and its injecting method.}
      The proposed attention-based injection can benefit from normal information without any side effects.
      }
      \label{fig:normal_ablation}
\end{figure}

\begin{figure}[t]
      \centering
      \includegraphics[width=\linewidth]{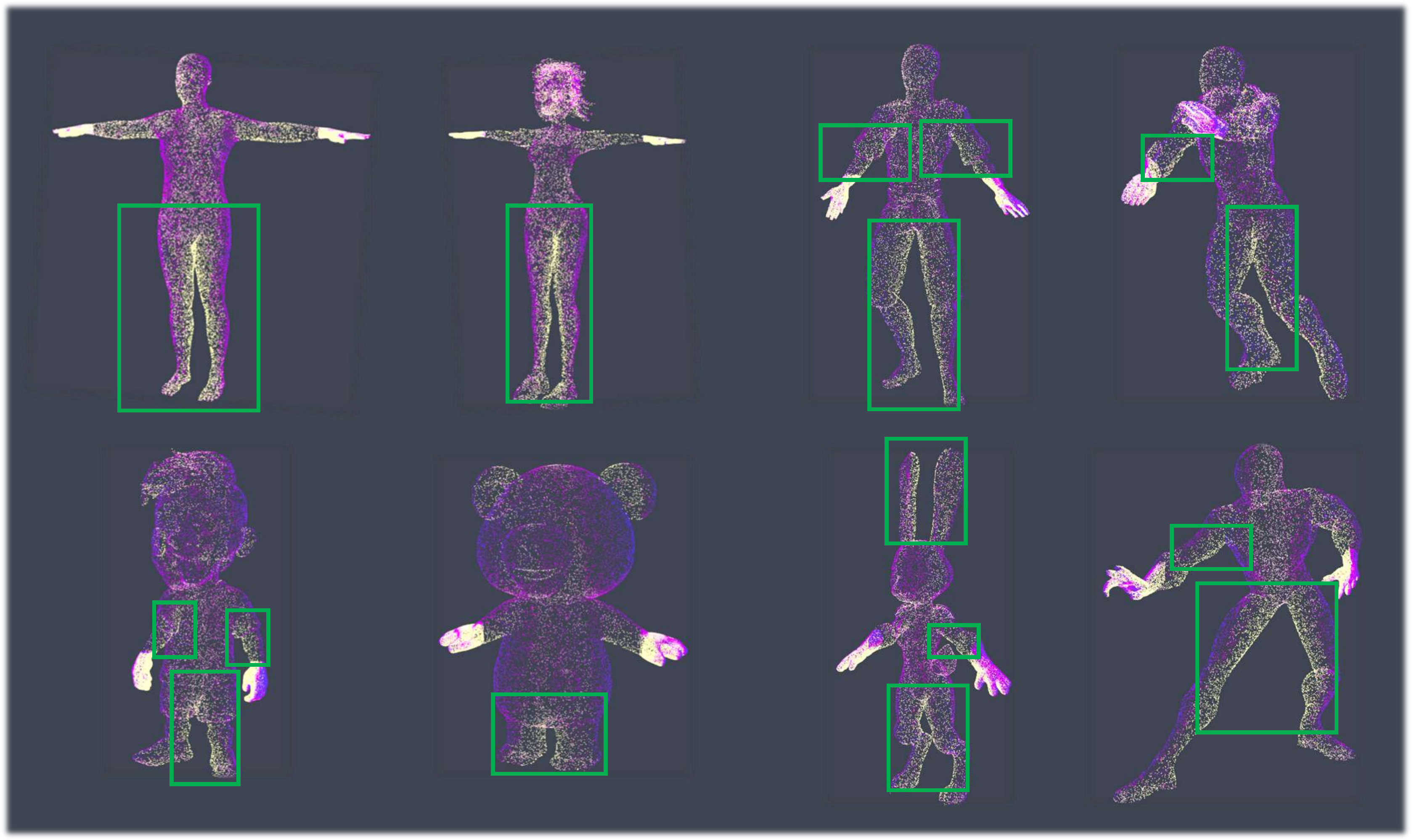}
      \caption{
      \textbf{Visualization of the attention score of our geometry-aware attention module.}
      These per-sampled-point values are extracted from the first attention head (out of 8 heads in total).
      The brighter color (yellower) indicates more attention to normals rather than coordinates.
      We also use green bounding boxes to label some clusters where high-attention-score points are densely distributed.
      It can be observed that the module adaptively learns to rely more on normals in regions like the inner thigh since coordinates become less discriminative there.
      }
      \label{fig:normal_attenion_score}
\end{figure}

\begin{figure*}[t]
      \centering
      \includegraphics[width=\linewidth]{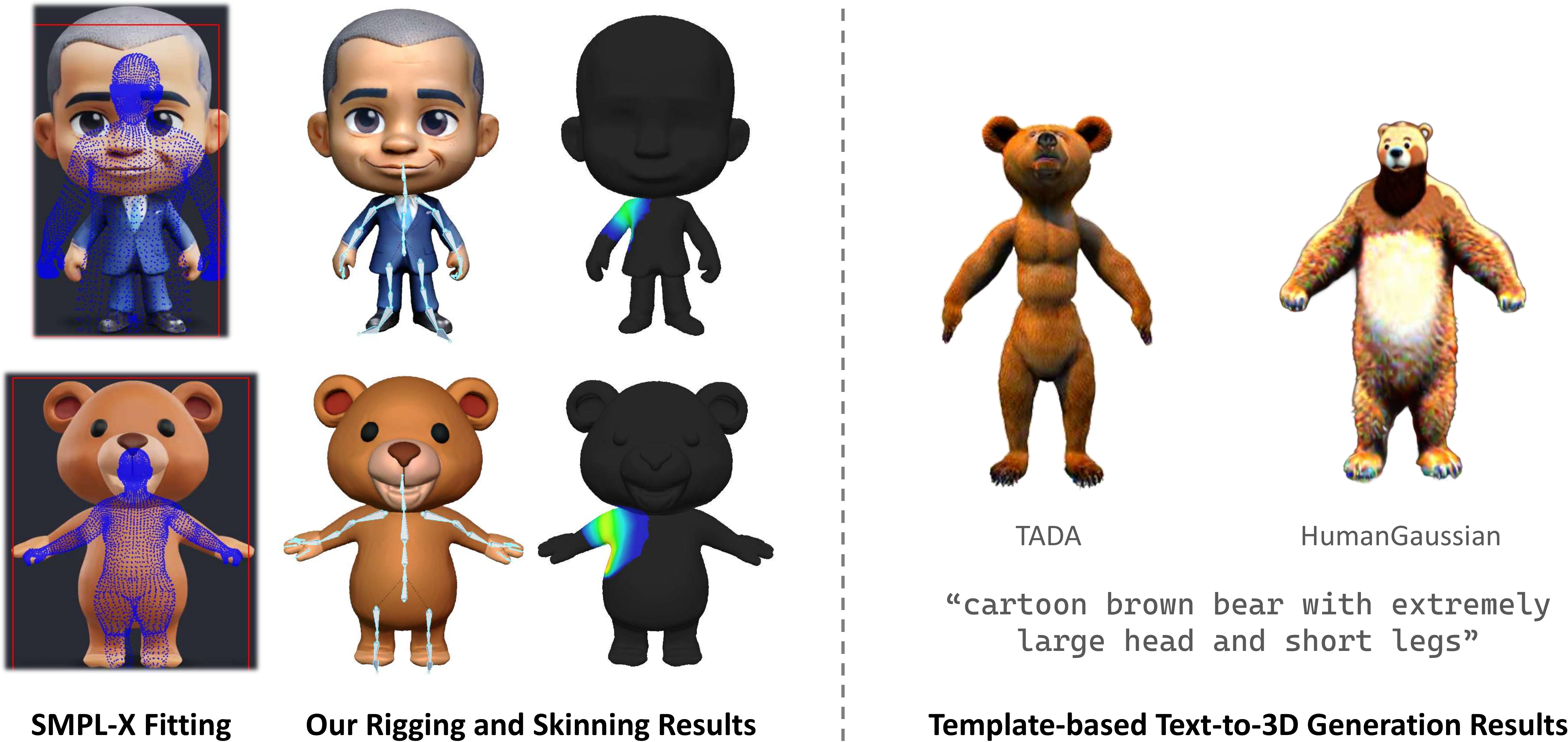}
      \caption{
      \textbf{Illustration of the limitations of SMPL-based rigging.}
      While SMPL provides a good template for skeletons and weights, it lacks the flexibility to handle exaggerated body shapes.
      }
      \label{fig:smpl}
\end{figure*}

\parsection{Tricky cases}
We show more results produced by our method, focusing on the details and some tricky cases. Each sub-figure of \cref{fig:tricky} proves some advantages of our methods, which is interpreted one by one as follows.

\parsection{(a)~Fine-grained control of fingers}
Thanks to the coarse-to-fine shape representation (\cref{sec:sampling}), our method shows remarkable accuracy at the hand regions, which is difficult for most existing approaches.

\parsection{(b)~Capacity of unusual shapes}
For those characters with an exaggerated body ratio (\eg, extremely large head, short limbs, etc), our method can adaptively change the bone length to fit the shape. This is intractable for template-based human-generating works.

\parsection{(c)~Complex input poses}
Poses far from the T-/A-pose are fully supported. Our model can not only produce the well-fitted posed skeleton, but can also transform it into the T-pose for further animating applications.

\parsection{(d)~Efficiency for high polygon models}
Benefiting from the particle-based shape autoencoder (\cref{sec:autoencoder}), our framework is efficient and robust for different input resolutions, ranging from low-poly meshes like (c) to practical game-level 3D models like (d). The Wukong model in (d) has over 1 million triangular faces, but our method can still make it animatable within 3 seconds.

\parsection{(e)~Support for asymmetric inputs}
While many rigging methods assume symmetric inputs, our method can effectively deal with asymmetric ones. For example, the big gun of this cyborg in (e) is bound well to his arm bones.

\parsection{(f)~Adaptation to non-existing bones}
This armless statue is predicted to have extracorporeal arm bones. However, due to the spatial query mechanism of our framework, those dummy bones have no weights toward the actual vertices and can be simply removed without any side effects.

\parsection{(g) \& (h)~Extension to extra bones}
Once fine-tuned with some extra data (\cref{sec:train}), our model is capable of producing positions and weights of non-standard bones. Here we show the long tail of a monkey in (g) and the long ears of a bunny in (h), which are all fully animatable.

\parsection{Geometry-awareness: attention vs. adding}
Intuitively, simply concatenating or adding the normals to the coordinates when feeding points into the shape encoder can achieve the same goal as our geometry-aware attention. However, we found in practice that such a vanilla injecting strategy often leads to overfitting on the high-quality training mesh normals. The weight prediction depends so much on normals that some regions are influenced by faraway bones just because they have plausible normals, especially when faced with lower-quality meshes (\eg, produced by generative models).
We present a typical case in \cref{fig:normal_ablation}. While injecting normal information via adding resolves the weight corruption problem in the armpit region, it introduces a new issue of incorrect weights on hands.
In contrast, the proposed attention mechanism benefits from normal information without any side effects.

\parsection{Geometry-aware attention score}
In addition to the two exemplar cases in \cref{fig:ablation}, here we visualize the geometry-aware attention score of more characters in \cref{fig:normal_attenion_score}.
The distribution of high-attention-score points shows patterns with statistical significance. Specifically, regions with possible spatial ambiguity, \eg, inner thigh, armpit, between ears, \etc, are affected more by the normal values.
This verifies the effectiveness of our geometry-aware attention, as it works exactly in the desired way to adaptively exploit the normal information.

\parsection{Limitations of SMPL-based rigging}
Template-based avatar generation methods~\cite{liao2024tada,liu2024humangaussian} benefit a lot from the well-defined SMPL mesh~\cite{loper2015smpl,pavlakos2019smplx}, enabling the generation of animatable characters with no additional rigging cost.
However, as discussed in \cref{sec:intro,sec:exp_results}, one of the unneglectable limitations of these methods is their inability to depart from realistic human shapes.
Although TADA~\cite{liao2024tada} attempts to address this issue by predicting vertex deformations of the template mesh, it remains constrained by the preset body ratio of SMPL.
\cref{fig:smpl} demonstrates some practical examples of cartoon characters with exaggerated body shapes that the SMPL model can hardly accommodate. As illustrated in the left part of \cref{fig:smpl}, the large heads of these characters cannot be fitted by the template meshes, and the interpolated blend weights will definitely be incorrect in the head regions.
In the right part of \cref{fig:smpl}, we exhibit the results generated by two template-based methods using the same text prompt ``cartoon brown bear with extremely large head and short legs''.
Despite the specific prompting, both methods still produce body ratios resembling those of realistic humans, which limits their applicability in practical scenarios.
In contrast, our method offers a promising solution by making bipedal characters of any shape ready for animation.

\begin{figure*}[t]
      \centering
      \includegraphics[width=\linewidth]{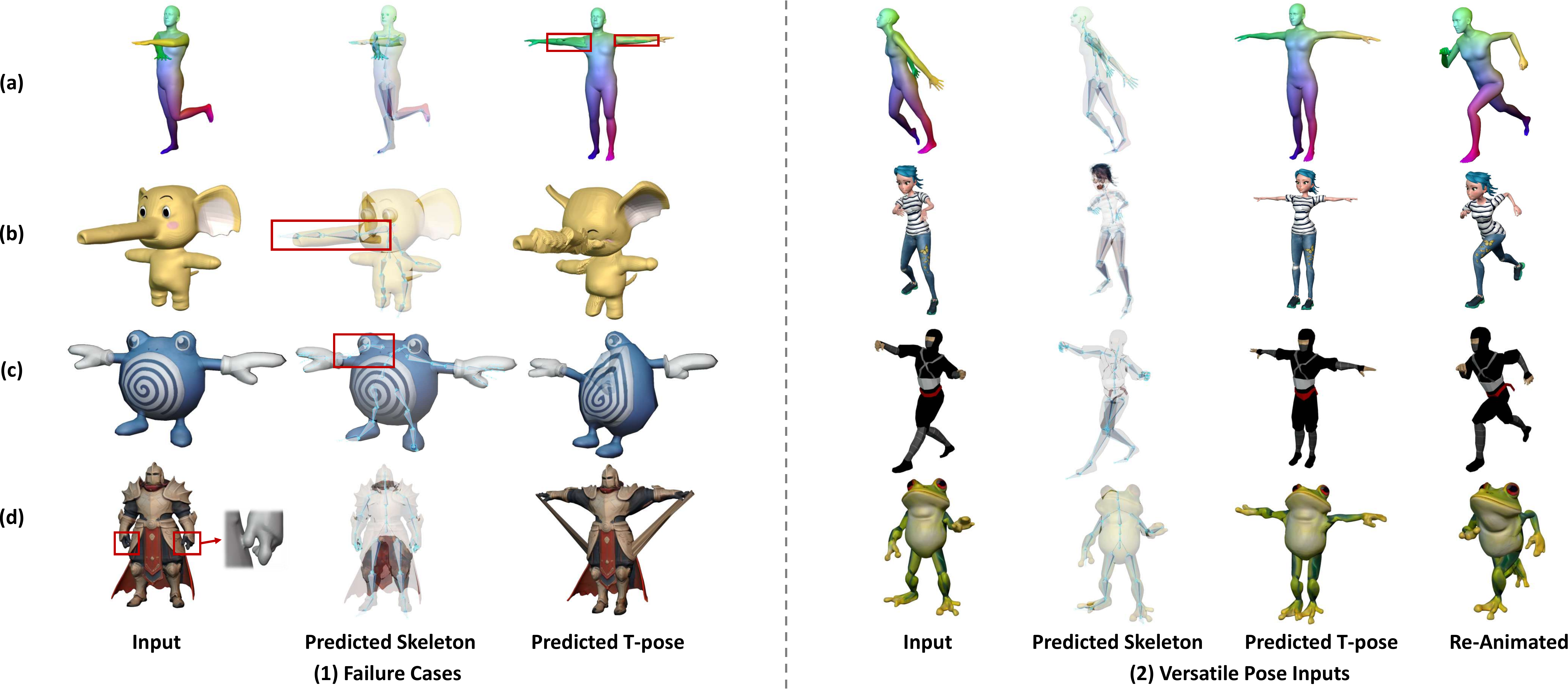}
      \caption{
      \textbf{(1) Failure cases of our method.}
      (a) Challenging pose with self-intersection; (b) \& (c) Out-of-distribution shapes; (d) Topological defects (unexpected mesh connections).
      \textbf{(2) More results produced by our method with non-rest-pose inputs.}
      }
      \label{fig:failure_poses}
\end{figure*}

\parsection{More versatile pose inputs}
It is worth noting that many of our evaluations above are based on inputs close to rest poses because the baselines are not specifically designed to handle versatile pose inputs.
Specifically,
RigNet~\cite{xu2020rignet} and TARig~\cite{ma2023tarig} cannot transform arbitrary poses into rest poses. Mixamo~\cite{mixamo} and Anything World~\cite{anythingworld} raise failure without inputs close to A- or T-pose. Meshy~\cite{meshy} and Tripo~\cite{tripo} perform poorly on non-rest poses (see \cref{fig:meshy_tripo} and \cref{fig:meshy_tripo_more}).
To show our method's generalizability of poses, we provide more results with non-rest inputs in \cref{fig:failure_poses}~(2).

\parsection{Failure cases}
\cref{fig:failure_poses}~(1) presents some failure cases of our method.
Limitations involving challenging poses and out-of-distribution shapes could be resolved by including more targeted training data.
As for the topological defects of input meshes, incorporating other methods to refine the 3D geometry might be a promising solution.

\ifarxiv\else

\section{Videos for Dynamic Visualization and Practical Applications}

For better visualization, we provide a video file as part of the supplementary material, which includes the following content:
\begin{itemize}
    \item An illustration of our inference pipeline and the data flow.
    \item The showcase of 3D characters and their animating results enabled by our framework. Inputs represented by mesh and 3D Gaussian Splats are both included.
    \item The process of using our web demo to make 3D characters animatable with one click.
    \item A real-world application that brings a character figure to live by integrating our framework with an off-the-shelf image-to-3D generator.
\end{itemize}

\fi

\end{document}